\documentclass[aps,twocolumn,groupedaddress,nofootinbib,english,nobalancelastpage,floatfix,superscriptaddress,showpacs]{revtex4-1}

\usepackage[latin1]{inputenc}
\pdfoutput=1
\usepackage{bm}
\usepackage{multirow, amsmath, amssymb, amsfonts, array}
\usepackage{color}
\usepackage{graphicx}
\usepackage{calc}
\usepackage{dcolumn}
\usepackage{mathrsfs}
\usepackage[normalem]{ulem}
\usepackage{cancel}

\def\bvec#1{\textrm{\boldmath $#1 $}}

\def\GeV{{\rm \ GeV}}
\def\MeV{{\rm \ MeV}}

\def\TeV{{\rm \ TeV}}

\def\der{{\rm \ d}}

\def\beq{\begin{equation}}
\def\eeq{\end{equation}}
\def\bea{\begin{eqnarray}}
\def\eea{\end{eqnarray}}
\def\simlt{\lower.5ex\hbox{$\; \buildrel < \over \sim \;$}}
\def\simgt{\lower.5ex\hbox{$\; \buildrel > \over \sim \;$}}
\def\simpropto{\lower.2ex\hbox{$\; \buildrel \propto \over \sim \;$}}

\def\apjs{ApJS}
\def\apj{ApJ}

\def\aap{Astronomy and Astrophysics}

\begin{document}

\title{Enhanced Line Signals from Annihilating Kaluza-Klein Dark Matter}

\author{Chiara Arina}
\email{Carina@uva.nl}
\affiliation{Institut d'Astrophysique de Paris - 98 bis boulevard Arago - 75014 Paris, France}
\affiliation{GRAPPA Institute, University of Amsterdam, Science Park 904, 1090 GL Amsterdam, Netherlands}

\author{Torsten Bringmann}
\email{Torsten.Bringmann@fys.uio.no}
\affiliation{Department of Physics, University of Oslo, Box 1048 NO-0316 Oslo, Norway}

\author{Joseph Silk}
\email{Silk@astro.ox.ac.uk}
\affiliation{Institut d'Astrophysique de Paris - 98 bis boulevard Arago - 75014 Paris, France}
\affiliation{1 Beecroft Institute of Particle Astrophysics and Cosmology, Department of Physics,
University of Oxford, Denys Wilkinson Building, 1 Keble Road, Oxford OX1 3RH, UK}
\affiliation{The Johns Hopkins University, Department of Physics and Astronomy,
3400 N. Charles Street, Baltimore, Maryland 21218, USA}

\author{Martin Vollmann}
\email{Martin.Vollmann@desy.de}
\affiliation{II. Institut f\"ur Theoretische Physik, Universit\"at Hamburg, Luruper Chaussee 149, 22761 Hamburg, Germany}

\begin{abstract}
Monochromatic gamma ray lines have long been known to provide potential smoking gun 
signals for annihilating dark matter. Here, we demonstrate that the situation is particularly 
interesting for Kaluza-Klein dark matter because resonant annihilation is generically expected 
for small, but not necessarily vanishing relative velocities of the annihilating particles. We 
calculate the contribution from those hitherto neglected resonances and show that the 
annihilation rate into monochromatic photons can be significantly enhanced, in a way that is 
much more pronounced than for the associated production of continuum photons. For favorable 
astrophysical conditions, this  leads to promising prospects for the detection of TeV-scale 
Kaluza-Klein dark matter. We also point out that the situation may be even more interesting in 
the vicinity of black holes, like the supermassive black hole at the 
center of our Galaxy, where in principle center-of-mass energies much larger than the rest mass 
are available. In this case, annihilating Kaluza-Klein dark matter may show the striking and 
unique signature of \emph{several} gamma ray lines, with an equidistant spacing corresponding 
to twice the compactification radius of the extra dimension.
\end{abstract}


\maketitle

%%%%%%%%%%%%%%%%%%%%%%%%%%%%%%%%%%%%%%%%%%%%%%%%%%%%%%%%%%%%%%%%%%%%%%%%%
\section{Introduction}

There is firm evidence for a sizable amount of dark matter (DM) on both cosmological and 
Galactic scales. While its nature still remains unknown, weakly interacting massive particles 
(WIMPs) are expected to be thermally produced in the early universe and thus  represent a 
theoretically very appealing class of possible 
candidates~\cite{Jungman:1995df,Bergstrom:2000pn,Bertone:2004pz}. 
Gamma rays provide a particularly promising means of indirectly searching for 
DM~\cite{Bringmann:2012ez}, not the least because the spectra from annihilating WIMPs often 
show prominent  features that do not only help significantly to distinguish DM signals from 
astrophysical backgrounds~\cite{Bringmann:2011ye} but also carry important information about 
the underlying particle physics model. The most prominent spectral feature is that of a 
monochromatic line from the loop-suppressed direct annihilation into 
photons~\cite{Bergstrom:1997fj}, while leading-order radiative corrections can produce sharp 
steps~\cite{Birkedal:2005ep,Bergstrom:2004cy} 
or somewhat broadened line-like structures~\cite{Bringmann:2007nk}; cascading decays of 
annihilation products might also give rise to box-shaped spectral features~\cite{Ibarra:2012dw}.

In view of the small galactic velocities, $v\sim10^{-3}$, the typically adopted approach in 
computing DM annihilation rates in this context is to take the $v\rightarrow 0$ limit such that the 
center-of-mass system (CMS) energy in collisions equals exactly twice the DM particles' mass. 
This greatly simplifies  otherwise rather intricate analytic calculations, both at the level of 
kinematics and amplitudes, and proved to be very useful when deriving  full 
one-loop~\cite{Bergstrom:1997fh,Bern:1997ng,Ullio:1997ke,Bergstrom:2004nr} or three-body 
final state~\cite{Bringmann:2007nk, Bringmann:2013oja} 
results for annihilation rates connected to the spectral features described above. 

{\it In this article, we consider situations where non-zero relative velocities lead to a significant 
enhancement of monochromatic gamma ray signals from annihilating DM or even introduce a 
new type of smoking-gun spectral signature}. In theories with additional spatial 
dimensions~\cite{Nordstrom:1988fi,Kaluza:1921tu,Klein:1926tv}, in particular, standard 
model (SM) fields are accompanied by a whole `tower' of heavy Kaluza-Klein (KK) states in the 
effective four-dimensional theory, with masses that are given by integer multiples of the inverse 
compactification scale \cite{Bailin:1987jd}. In such setups, the lightest new state may constitute 
a viable DM candidate and one inevitably expects to encounter resonances in the annihilation 
rate, due to the presence of other KK particles with almost exactly twice the DM particles mass. 
In fact, in astrophysical environments where DM collisions with very large CMS energies are 
possible, resonances with larger even-integer multiples of the DM mass could be reached -- with 
the exciting prospect of producing, in principle, a comb-like structure of equidistant gamma ray 
lines. Such a signature would, if observed, not only be a smoking-gun signature of particle DM 
but also unequivocally point to its underlying extra-dimensional origin.

While the situation sketched above is generic to KK DM models, we will in the following mostly 
restrict our discussion, for definiteness, to the case of one universal extra dimension (UED)~
\cite{Appelquist:2000nn} in which all SM fields can propagate. In this case, the first KK excitation 
of the photon is a viable DM 
candidate~\cite{Cheng:2002iz,Belyaev:2012ai,Servant:2002aq,Cheng:2002ej} that 
has been intensely discussed in the literature -- both in terms of prospects for accelerator 
searches~\cite{Datta:2005zs,Chen:2009gz,Cacciapaglia:2013wha,Servant:2014lqa} 
and direct DM 
detection~\cite{Servant:2002hb,Bertone:2007xj,Arrenberg:2008wy,Arrenberg:2013paa}, 
as well as indirect DM searches with 
gamma rays~\cite{Bertone:2002ms,Bergstrom:2004cy,Bergstrom:2004nr,Bertone:2010fn}, 
positrons~\cite{Hooper:2004xn,Baltz:2004ie,Hooper:2009fj}, 
neutrinos~\cite{Hooper:2002gs, Flacke:2009eu,Blennow:2009ag,Bertone:2010ww}
 or antiprotons~\cite{Bringmann:2005pp,Barrau:2005au} (for a review, see 
 Ref.~\cite{Hooper:2007qk}). The relevance of resonances for the DM phenomenology of this 
 model~\cite{Kakizaki:2005en} has been explored in detail for both collider 
 searches~\cite{Battaglia:2005zf,Bhattacherjee:2005qe,Matsumoto:2009tb} and precision 
 computations of the relic 
 density~\cite{Kakizaki:2005uy,Kakizaki:2006dz,Matsumoto:2009tb,Cornell:2014jza}. In 
 this article, we demonstrate that resonances can be at least as 
 important for indirect detection, and calculate annihilation rates for processes that produce 
 pronounced spectral signatures in gamma rays.

We study various situations of interest where the full velocity-dependence of the annihilation 
rate must be taken into account. For DM annihilation in the Galactic halo, e.g., averaging over 
the DM velocity distribution can significantly enhance the annihilation rate into monochromatic 
photons, compared to the $v\rightarrow0$ limit, even for small average velocities. A 
supermassive black  holes (SMBH) like the one at the Galactic center (GC), on the other hand, 
can act as an effective DM particle accelerator with CMS energies many times above the rest 
mass~\cite{Baushev:2008yz,Banados:2009pr}. Whether the effect of those high-energy 
collisions close to the horizon is actually observable far away from the black hole 
(BH)~\cite{Banados:2010kn,Williams:2011uz,Cannoni:2012rv,Grib:2012iq,Patil:2012fu,Harada:2012ap}, 
or rather not~\cite{Jacobson:2009zg,Berti:2009bk,McWilliams:2012nx}, is a matter of ongoing 
debate. Here, we point out that if the annihilation rate is sufficiently large, upcoming GC 
observations at TeV energies by HESS~\cite{Aharonian:2006pe} or 
CTA~\cite{Acharya:2013sxa} may indeed  reveal the unique and striking spectral signature of a 
line `forest'  that we have already mentioned above.

This article is organised as follows. We start by introducing the UED model in 
Sec.~\ref{sec:ued}, 
with a particular focus on its minimal version.  The resulting gamma ray spectrum from 
KK DM is then addressed in Sec.~\ref{sec:sigma}, including in particular a detailed discussion of 
the hitherto neglected contributions from various resonances in the relevant annihilation cross
sections. Next, we introduce in Sec.~\ref{sec:astro} the astrophysical setup that is required to 
translate annihilation rates into the expected gamma ray flux, including a short discussion about 
DM halo density and velocity distributions. In Section~\ref{sec:le} we apply this formalism to our 
results from Sec.~\ref{sec:sigma} in order to assess  the impact of a non-vanishing, but still not 
highly relativistic, DM velocity on the gamma ray signal from KK DM. Section~\ref{sec:he} then 
focuses on the more optimistic case of highly relativistic DM particles accelerated by 
SMBHs -- which offers, as we will see, the possibility of a particularly striking 
signature in gamma rays. In Sec.~\ref{sec:concl}, finally, we summarize our results and 
conclude. The technical details on the calculation of the annihilation process are given in 
Appendices~\ref{app:KKdec},~\ref{app:KKann} and~\ref{app:h2kres}, while the details for the 
photon flux from the region around a BH can be found in Appendix~\ref{app:BHdet}.

%%%%%%%%%%%%%%%%%%%%%%%%%%%%%%%%%%%%%%%%%%%%%
%%%%%%%%%%%%%%%%%%%%%%%%%%%%%%%%%%%%%%%%%%%%%
\section{Universal extra dimensions}
\label{sec:ued}

The UED model is essentially a higher-dimensional version of the SM of particle physics, i.e.~all 
SM fields are allowed to propagate in one or more compactified extra 
dimensions~\cite{Appelquist:2000nn}. 
Since the SM in $d>4$ is not renormalizable, this must be understood as an effective field 
theory which is only valid up to a cutoff scale $\Lambda$. We will restrict our discussion in the 
following to the simplest case of one UED, where the extra dimension is compactified on an 
$\mathbb{S}^1/\mathbb{Z}_2$ orbifold. The orbifold construction is essential to both recover the 
chiral structure of the 4D effective theory, which is non-trivial in view of the fact that chiral 
fermions do not exist in 5D, and to prevent unwanted light degrees of freedom that correspond 
to the higher-dimensional components of gauge fields. 

5D fields, or rather all their components with a well-defined behavior under 4D Lorentz 
transformations,  can then be expanded as either
\beq
 \Phi(x^\mu,y)=\frac{1}{\sqrt{2\pi R}}\phi^{(0)}(x^\mu)+\frac{1}{\sqrt{\pi R}}\sum_{n=1}^\infty
 \phi^{(n)}(x^\mu)\cos\frac{ny}{R}\label{even}
\eeq
or
\beq
\Phi(x^\mu,y)=\frac{1}{\sqrt{\pi R}}\sum_{n=1}^\infty\phi^{(n)}(x^\mu)\sin\frac{ny}{R}\,,\label{odd}
\eeq
depending on whether one assigns even (\ref{even}) or odd (\ref{odd}) transformation properties 
under the orbifold projection $y\rightarrow-y$. Here and in the following, $x^\mu$ denote ordinary 
4D space-time co-ordinates, $y$ the extra dimensional direction and $R$ the compactification 
radius.  Each 5D $\Phi$ thus corresponds to a whole tower of heavy states $\phi^{(n)}$ in the 
effective 4D theory that is obtained after integrating out the extra dimension; only even states, 
however, have a light zero mode $\phi^{(0)}$ (which is identified with the corresponding SM 
field).

In this way, each SM gauge field $A_\mu$ is accompanied by a tower of KK states, 
$A_\mu^{(n)}$, and each fermion $\psi$ by {\it two} towers that represent $SU(2)$ doublets, 
$\psi_{\rm d}^{(n)}$, and singlets, $\psi_{\rm s}^{(n)}$, respectively. As for the scalar sector, there 
is the SM Higgs field $h$ and its KK tower $h^{(n)}$; further physical states $a^{(n)}$ and 
$a_\pm^{(n)}$ arise at 
$n\geq1$ as linear combinations of the higher-dimensional components  of the 
$SU(2)_L\times U(1)_Y$ gauge fields and would-be Goldstone bosons of the Higgs doublet. Up 
to the first KK-level, $n\leq1$, the spectrum of states is thus the same as in the minimal 
supersymmetric standard model, up to the spin properties of the SM partners, which is why the 
UED model has sometimes also  loosely been referred to as 
`bosonic supersymmetry'~\cite{Cheng:2002ab}.

As a direct consequence of momentum conservation along the extra dimension, KK 
number is conserved at tree level. Every tree-level vertex involving particles with KK number 
$n_i$ must thus obey one of the selection rules implied by $\sum \pm n_i=0$.
At higher orders in perturbation theory, on the other hand, this 
is not necessarily the case because the orbifold fixpoints break the original 5D translational 
invariance. One of the phenomenologically most important consequences of the orbifold 
compactification, however, is that KK-parity, defined as $(-1)^{\sum_i n_i}$, is still 
conserved.\footnote{
This can be traced back to the invariance of (\ref{even},\ref{odd}) under reflection about the 
center ($y=\pi R/2$) of the extra-dimensional interval.
} 
This remnant of the original 5D translational invariance implies that the lightest Kaluza-Klein 
particle (LKP) is stable and thus a potential DM candidate (in the very same fashion as the 
lightest supersymmetric particle is stable if $R$-parity is conserved).
For our discussion, however, the actual mass spectrum is not only  crucial for determining the 
LKP but also because it determines both the exact location of resonances and which decay 
processes are kinematically allowed. 

At tree level, the mass of a KK-state is given by
\beq
 \label{mtree}
  M_{(n)}^i= m_{\rm EW}^i+\left(\frac{n}{R}\right)^i\,,
\eeq
where $i=2$ ($i=1$) for bosons (fermions) and $m_{\rm EW}$ is the mass of the corresponding SM state, generated by electroweak 
symmetry breaking. Current collider data constrain the compactification scale to 
$R^{-1}\gtrsim700$\,GeV \cite{Edelhauser:2013lia,Kakuda:2013nva,Kakuda:2013kba}, with 
stronger limits applying for $d>5$. One thus generically expects very degenerate spectra at any 
given KK level, due to $R^{-1}\gg m_\text{EW}$,  which implies that radiative mass corrections 
$\delta M_{(n)}$ become important to determine the actual mass hierarchy of states. Those 
corrections on top of the SM contributions (which are renormalized in the usual way) arise both 
due to winding modes of loops in the bulk and due to terms localized at the orbifold boundaries 
\cite{Cheng:2002iz,vonGersdorff:2002as,Georgi:2000ks}. The latter are formally infinite and 
thus need to be renormalized by counter-terms with in general unknown finite parts. 
 
 The scenario of {\it minimal} UED (mUED) rests on the simplifying assumption 
 that those terms at the orbifold boundaries can be neglected at the cutoff scale $\Lambda$; all 
 KK masses are then uniquely determined by only two parameters $\Lambda$ and 
 $R$~\cite{Cheng:2002iz}. For the recently determined Higgs mass of 
 $m_h\sim 125$\,GeV \cite{Aad:2012tfa,Chatrchyan:2012ufa}, however, the running of the 
 Higgs self-coupling implies an 
 unstable vacuum unless the cutoff scale is as small as $\Lambda R\sim5$ 
 \cite{Blennow:2011tb,Kakuda:2013kba,Cornell:2014jza}. 
 In the mUED scenario, the LKP is the first KK excitation of the photon, 
 which to a very good approximation is the same as the first KK excitation of the hypercharge 
 gauge boson, $B^{(1)}$. As it turns out, the $B^{(1)}$ is indeed an excellent DM  
 candidate~\cite{Servant:2002aq}. Thermal production in the early universe leads to the correct 
 relic density for $R^{-1}\simeq m_{B^{(1)}}\sim1.2$\,TeV \cite{Belanger:2010yx,Cornell:2014jza}, 
 a compactification scale that may well be in reach for the LHC after its 
 upgrade~\cite{Belyaev:2012ai}. Taking into account the requirement of relic density and 
 vacuum stability, there are thus essentially no free parameters in the simplified UED scenario 
 known as mUED.
 
The interaction terms localized at the orbifold fixpoints, however, are in principle arbitrary 
(though  their scale-dependence  is determined by bulk interactions). At a given scale,  they 
should thus in general simply be viewed as new free parameters of the theory. In particular, 
those parameters should follow from some more fundamental theory at energies 
$E\gtrsim \Lambda$ and there is no obvious reason why such a theory should predict all those 
terms to vanish at the cutoff scale. Compared to  the mUED scenario, non-vanishing boundary 
terms will affect the corrections to the self-energies, and thus the mass hierarchy, of KK 
particles. In fact, in non-minimal UED scenarios, one may explicitly allow even for bulk mass 
terms~\cite{Flacke:2013pla}. 
An often adopted approach is therefore to treat the mass splittings $\delta M^{(n)}$ as 
essentially free parameters. We note that this could even change the nature of the 
LKP~\cite{Flacke:2008ne}, but leave 
an investigation of possible consequences of this interesting possibility for future work. 
Changing the mass-splittings of the KK particles can have significant effects on the 
compactification scale that results in the correct relic density for a $B^{(1)}$ LKP. The lowest 
possible value is given by $R^{-1}\sim800$\,GeV and corresponds to mass splittings much 
larger than in the mUED case, such that co-annihilations are no longer important 
\cite{Servant:2002aq}. Tuning the mass spectra to be highly degenerate, on the other hand, 
makes co-annihilations even more important and may drive the compactification scale, and thus 
mass, of a thermally produced $B^{(1)}$ LKP up to a value of a few TeV 
\cite{Kong:2005hn,Burnell:2005hm}. Finally, let us stress that the cutoff scale will in general be 
significantly larger than the value of $\Lambda R\sim5$ implied by vacuum stability in the 
simplified mUED scenario.

%%%%%%%%%%%%%%%%%%%%%%%%%%%%%%%%%%%%%%%%%%%%%
%%%%%%%%%%%%%%%%%%%%%%%%%%%%%%%%%%%%%%%%%%%%%
\section{Photons from KK dark matter}
\label{sec:sigma}

%%%%%%%%%%%%%%%%%%%%%%%%%%%%%%%%%%%%%%%%%%%%%
\subsection{Spectrum in the zero-velocity limit}
\label{sec:sigold}

The expected gamma ray spectrum from the annihilation of $B^{(1)}$ pairs to SM model 
particles has been extensively studied in the literature. First of all, there is the usual {\it 
secondary} contribution to the spectrum  from $B^{(1)}B^{(1)}\rightarrow \bar q q,ZZ,W^+W^-$ 
\cite{Bertone:2002ms} that 
results from the fragmentation and decay of the annihilation products, mostly via 
$\pi^0\rightarrow\gamma\gamma$. Unlike the typical situation in supersymmetry, also the decay 
of $\tau$ leptons gives an important contribution and leads to a significantly harder spectrum 
\cite{Bergstrom:2004cy} as a result of the relatively large $B^{(1)}B^{(1)}\rightarrow\tau^+\tau^-$ 
rate.

In fact, the annihilation into lepton final states is the dominant channel, with roughly the same 
branching fraction of $\sim$20\% for all lepton families.  An even more important contribution at 
the phenomenologically most relevant highest energies, i.e.~close to the kinematical endpoint of 
$E_\gamma=m_\chi$, are thus {\it primary} photons radiated off lepton final legs 
\cite{Bergstrom:2004cy}. This final state radiation (FSR) is dominated by collinearly emitted 
photons, resulting in a universal spectrum of the Weizs\"acker-Williams 
form~\cite{Birkedal:2005ep,Bergstrom:2004cy,Beacom:2004pe}:
\bea
\frac{dN^{\rm FSR}_\gamma}{dx}&\equiv&
\frac{1}{ \sigma_{B^{(1)}B^{(1)}\rightarrow\ell^+\ell^-}}\frac{d\sigma_{B^{(1)}B^{(1)}
\rightarrow\ell^+\ell^-\gamma}}{dx}\\
  &\simeq& 
  \frac{\alpha}{\pi}\frac{1+(1-x)^2}{x}\log\left(\frac{s(1-x)}{m_\ell^2}\right)\,.\label{ww}
\eea
Here, $\sqrt s=2m_{B^{(1)}}$ is the CMS energy and 
$x\equiv2E_\gamma/\sqrt{s}=E_\gamma/m_{B^{(1)}}$. Overall, one expects  a characteristic, 
relatively hard spectrum which drops abruptly at the DM mass; such a photon distribution could 
very efficiently be discriminated from typical astrophysical backgrounds \cite{Bringmann:2011ye}. 

An even more striking spectral feature would be the quasi-monochromatic line expected for 
$B^{(1)}B^{(1)}\rightarrow \gamma X$, at a photon energy of
\beq
\label{eline}
E_\gamma=m_{B^{(1)}}\left(1-\frac{m_X^2}{4m_{B^{(1)}}^2}\right)\,. 
\eeq
Due to the large LKP mass, the three possible line signals (for $X=\gamma,Z,h$) would 
essentially be indistinguishable and thus simply add up in the spectrum 
(at $E_\gamma\simeq m_\chi$). A fully analytic one-loop calculation has been performed for the 
dominant process of  $B^{(1)}B^{(1)}\rightarrow \gamma \gamma$ via fermion box 
diagrams~\cite{Bergstrom:2004nr}. Numerical calculations have both confirmed and extended 
these analytic results \cite{Bertone:2010fn}, as well as estimates  \cite{Bergstrom:2004nr} for 
the subdominant annihilation channels into $\gamma Z$ and $\gamma h$ final states. In order 
to discriminate the monochromatic signal from the continuum FSR photon signal discussed 
above, given an expected total annihilation cross section of 
$(\sigma v)_{\gamma X} \lesssim 10^{-29}\, \rm cm^3/s$, requires the energy resolution of the 
detector to be better than a few percent~\cite{Bergstrom:2004nr}. 
While such a performance is, unfortunately, unfeasible for both operating and  
upcoming Air Cherenkov Telescopes, which feature energy resolutions of 10-15\%, it might be 
well in reach for  space-based telescopes given the design characteristics of planned missions 
like Gamma-400~\cite{Galper:2012ji}, DAMPE~\cite{dampe} or CALET~\cite{Mori:2013ida}.  

It is worth stressing that the continuous gamma ray spectrum from annihilating $B^{(1)}$ pairs is 
rather  insensitive to the other KK masses, such that one expects essentially the same spectrum 
even in non-minimal UED scenarios. The strength of the line signal, on the other hand, can be 
enhanced by a factor of a few when allowing for smaller mass differences between KK fermions 
and the $B^{(1)}$~\cite{Bergstrom:2004nr} (and can be much larger for other LKP candidates, 
such as the $Z^{(1)}$~\cite{Bonnevier:2011km}).

%%%%%%%%%%%%%%%%%%%%%%%%%%%%%%%%%%%%%%%%%%%%%
\subsection{Annihilation rate revisited}
\label{sec:signew}

Let us now address the question of how the above presented situation changes when allowing 
for  a non-zero relative velocity of the annihilating LKP pair. The first thing to note is that the 
FSR continuum spectrum $dN^{\rm FSR}/dx$ will not change visibly if, as already indicated in 
Eq.~(\ref{ww}), one uses the actual CMS energy rather than $2m_{B^{(1)}}$ in  defining the 
dimension-less photon energy $x$. The same is true for the secondary photons, given that $s$ 
is the only scale in the problem (provided that, as is the case of interest here, the CMS energy is 
much larger than the mass of any of the annihilation products). Unless one is in the highly 
relativistic regime, furthermore, one can expect even the normalization of the spectrum to stay 
roughly constant because the $B^{(1)}$ is an $s$-wave annihilator with a total annihilation cross 
section of 
\beq
  \sigma v_{\rm rel}\simeq3\times10^{-26}{\rm cm}^3{\rm s}^{-1} 
  \left(\frac{m_{B^{(1)}}}{800\,{\rm GeV}}\right)^{-2}\,,
  \label{eq:scaling}
\eeq
where both $\sigma v$ and the final state branching ratios are rather insensitive to 
the spectrum of other KK states \cite{Bringmann:2005pp}. The same expectations hold for the 
line signals discussed above: while the location of the line will shift from 
$E_\gamma\simeq m_\chi$ to $E_\gamma\simeq \sqrt{s}/2$, its normalization will stay roughly 
the same as long as the CMS energy is not significantly larger than the rest mass of the two 
annihilating LKPs.

There is one important exception to these considerations and this is what we will focus on in the 
following: the appearance of $s$-channel {\it resonances} may significantly enhance the 
annihilation rate with respect to the zero velocity limit (in which case $s$-channel diagrams give 
subdominant contributions for both the line and continuum signals).
A further advantage of these resonances is that they add a scale to the process, which in 
general is the only way to preserve a sharp spectral feature in the potentially observable 
gamma ray flux after integrating over a {\it distribution} of CMS energies or relative velocities 
(see also Section \ref{sec:astro}).\footnote{
This is most easily seen for the case of a monochromatic line: integrating 
$d\sigma/dE_\gamma\equiv N\,\delta(E_\gamma-\sqrt{s}/2)$ over some -- astrophysically 
motivated and  typically featureless -- CMS energy distribution $f(s)$ simply results in a flux 
proportional to $f(2E_\gamma)$ for an energy-independent normalization $N$; the initial 
line-feature is thus completely smeared out. If, on the other hand, $N$ is strongly peaked, at the 
energy of the resonance, a pronounced  peak at the same energy will also show up in the flux -- 
independently of the functional form of $f$.
}

For at least three reasons, these observations are particularly relevant for the case of KK DM: 
\begin{enumerate}
\item
Due to the mass degeneracy of KK states, relevant  
resonances are naturally expected for level-2 KK states in the $s$-channel.
\item
 The decay of 
these resonances into SM states is necessarily loop-suppressed because it violates KK number 
conservation. This implies very narrow widths, and thus large enhancements on resonance, if the decay to level-1 KK 
states is kinematically forbidden or otherwise suppressed (which, as discussed below, often is the case). 
\item
Another consequence of a loop-suppressed total width is that continuum and monochromatic photons are produced at roughly the same strength on resonance, unlike the typical situation where only the line signal is loop-suppressed. In other words, one can expect a 
much larger {\it relative} enhancement of the line signals (which, as discussed above, is not the least needed to overcome the large contribution from FSR photons). 
\end{enumerate}

\begin{quote}
{\it In the UED scenario, resonances thus indeed single out spectral features in a 
unique way}. 
\end{quote}

\begin{figure}[t]
\begin{center}
  \includegraphics[width=\columnwidth,clip]{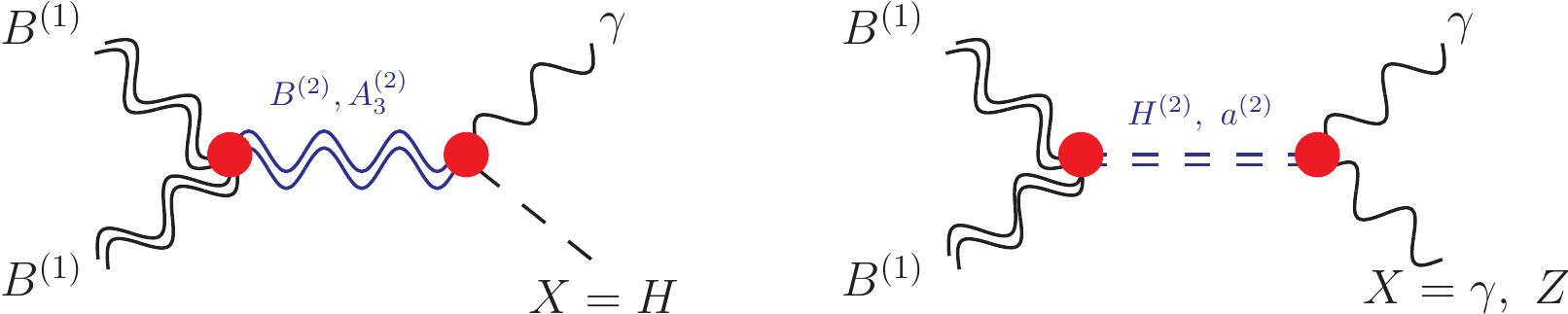}
\end{center}
\caption{Diagrams that generally lead to the most pronounced spectral features in the UED 
scenario when allowing for relative WIMP velocities $v\neq0$. The blobs correspond to effective
couplings that are computed in Appendices~\ref{app:KKdec} and \ref{app:KKann}. 
\label{fig:feynbbtogammax}} 
\end{figure}

\noindent
With these general considerations in mind, let us now turn to a more detailed discussion of 
which resonances will be most relevant in our case. Charge conservation implies that for the 
annihilation of a $B^{(1)}$ pair the only possible resonances at KK-level 2 are the vector bosons 
$B^{(2)}$, $A_3^{(2)}$ and the scalars $H^{(2)}$, $a_0^{(2)}$. In Fig.~\ref{fig:feynbbtogammax}, 
we show the corresponding Feynman diagrams. Here, the blobs on the left represent effective 
$B^{(1)}B^{(1)}Y^{(2)}$ couplings that may either exist at tree level or correspond to 1-loop sub-
diagrams. The right blob represents a KK-number violating coupling and is thus necessarily 
loop-suppressed.
However, not all combinations of resonance states $Y^{(2)}$ and final states $\gamma X$ are 
actually possible. For 
a scalar resonance, for instance, $X$ must  be a vector in order to 
conserve helicity. Vector resonances, on the other hand,  are only allowed for 
$X= H$: the $\gamma\gamma$ annihilation channel is forbidden by the Landau-Yang 
theorem \cite{Landau:1948kw,Yang:1950rg};   $\gamma Z$ final states cannot appear due 
to the anomaly cancellation familiar from the SM, which prevents anomalous three-gauge-boson 
couplings. For a very similar reason, in fact, it turns out that the $a_0^{(2)}$ resonance cannot 
decay into two vectors either (recall that $a_0$ contains the fifth component of the 
higher-dimensional $Z$ boson).

 \begin{figure}[t]
\begin{center}
  \includegraphics[width=\columnwidth,clip]{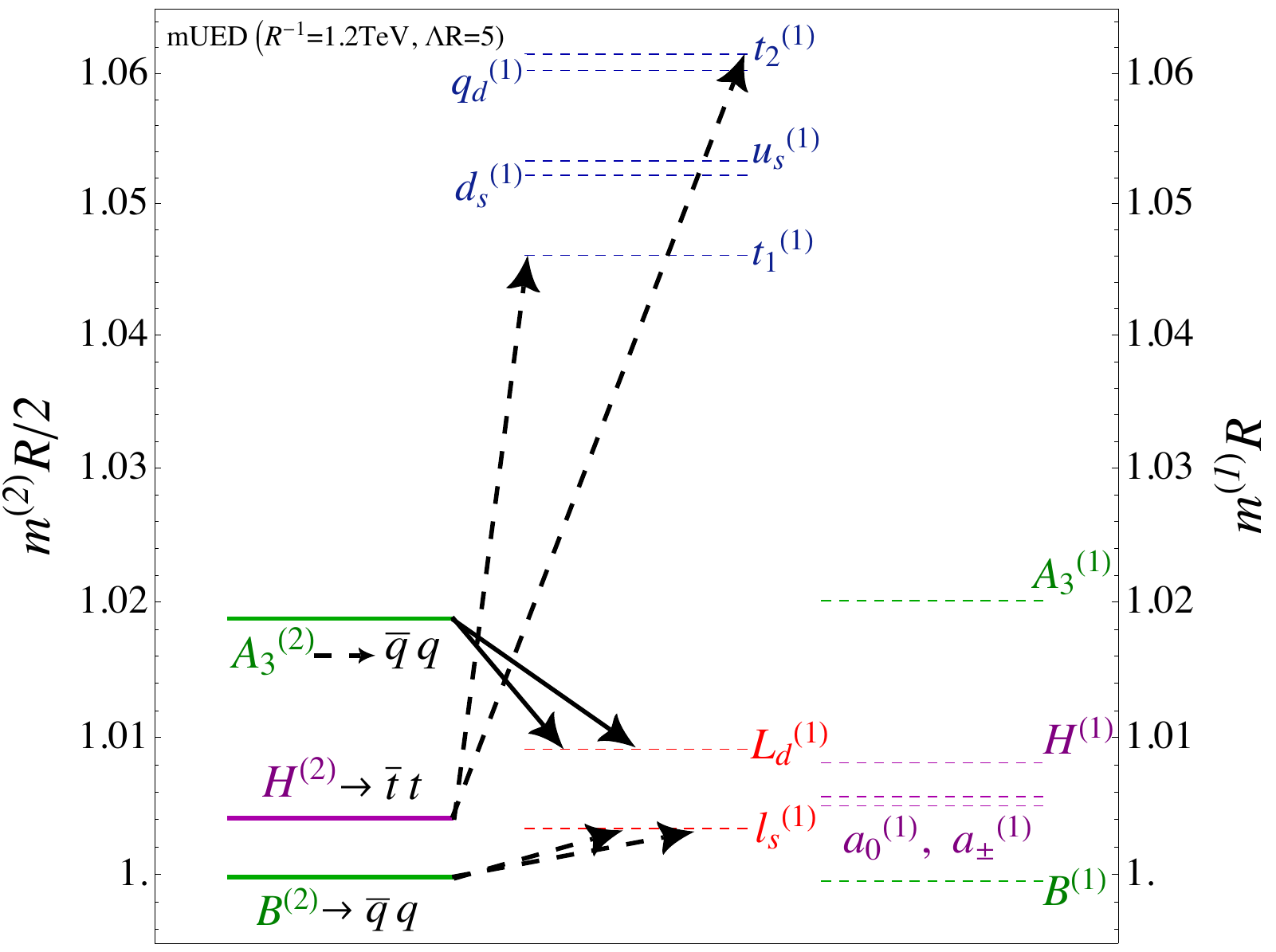}
\end{center}
\caption{Mass spectrum of relevant  KK(2) resonances in the mUED scenario, in units of  
{\it twice} the inverse compactification scale $R^{-1}$ (left column). The middle and right 
columns show  the mass of KK(1) states; note that only for the KK top quark the mass 
eigenstates $(t_1^{(1)},t_2^{(1)})$ differ significantly from the flavor eigenstates 
$(t_s^{(1)},t_d^{(1)}$). 
Dominant decay channels in the mUED case are displayed by solid arrows unless the  
resonance mainly decays to SM particles.  Dashed 
arrows indicate the dominant decay process in non-minimal UED versions. 
\label{fig:atomicphysplot}}
\end{figure}

 \begin{table}[t!]
\begin{tabular}{|c|c|c|c|}
\hline
&&&\\[-1.5ex]
{\bf Resonance $Y^{(2)}$} & \boldmath$B^{(2)}$ & \boldmath$A_3^{(2)}$ & \boldmath$H^{(2)}$\\[1ex]
\hline
\hline
&&&\\[-1.5ex]
\multirow{2}{*}{$Y^{(2)}\to \gamma X$} & $\gamma H$ & $\gamma H$ & $\gamma\gamma$, $\gamma Z$ \\[0.8ex]
 & \small$(\Gamma\!\sim\!0.07)$ & \small$(\Gamma\!\sim\!0.07)$ &  \small$(\Gamma\!\sim\!0.12,0.36)$  \\[0.8ex]
\hline
&&&\\[-1.5ex]
 \multirow{2}{*}{$\Gamma^\textrm{main}_{Y^{(2)}}$ {\scriptsize (mUED)}} &  $\bar f_\text{SM} f_\text{SM}$ & $\bar l^{(1)}_dl^{(1)}_d$ & $\bar t t$ \\[0.8ex]
 & \small$(\Gamma\!\sim\!0.8)$ & \small$(\Gamma\!\sim\!70)$ & \small$(\Gamma\!\sim\!0.1)$  \\[0.8ex]
\hline
&&&\\[-1.5ex]
\multirow{2}{*}{$\Gamma^\textrm{main}_{Y^{(2)}}$  {\scriptsize (non-mUED)}} & $\bar f^{(1)}_{d,\ s} f^{(1)}_{d,\ s}$ & $\bar f_\text{SM}f_\text{SM}$ & $\bar t^{(1)}_{d,s} t^{(1)}_{s,d}$\\[0.8ex]
 &  \small$(\Gamma\!\sim\!15)$ & \small$(\Gamma\!\sim\!0.8)$ & \small$(\Gamma\!\sim\!160)$  \\[0.8ex]
\hline
&&&\\[-1.5ex]
$B^{(1)}B^{(1)}Y^{(2)}$  & $\sim g'^3m_t$ & $\sim g'^2g\,m_t$ & $\sim g'^2g^{-1} m_W$  \\[0.8ex]
\hline
\end{tabular}
\caption{\label{tab:resonances} Main decay channels, couplings and possible $\gamma X$ final 
states for the resonances shown in Fig.~\ref{fig:feynbbtogammax} (note that 
$\Gamma_{a_0^{(2)}\to\gamma\gamma,\gamma Z}=0$). Decay rates are given in GeV 
and obtained for $R^{-1}=1.2$\,TeV and $\Lambda R=5$; see Appendix~\ref{app:KKdec} for 
calculational details.}
\end{table}
% deltaM ~ 0.1TeV, 0.2TeV for m_B^(2)=2.6TeV

The obvious next step consists in identifying which of the remaining processes are most 
relevant in producing line signals. To do so, it is instructive to have a closer look at the actual 
mass spectrum of the involved states. In Fig.~\ref{fig:atomicphysplot}, we show in the left 
column the mass of the relevant resonant particles $Y^{(2)}$ (in units of {\it twice} the inverse 
compactification scale $R^{-1}$). 
For comparison, the middle and right column show the mass of first-level excitations. 
The first thing to note is that the tree-level decay of $Y^{(2)}$ into KK-1 states is in some cases 
not kinematically possible, or at least heavily suppressed. 
% H^2 -> B^1 B^1 some kinematic suppression, but mostly because of (mW/ mB1)^2
The decay width for those particles is 
therefore instead determined by the loop-suppressed decay into two SM particles; 
such a narrow width will correspondingly enhance the LKP annihilation rate on resonance. 
The dominant decay channels are shown in the figure and also summarized in 
Tab.~\ref{tab:resonances}.  For comparison, we also indicate how this 
would change if all final states were kinematically accessible, as can be arranged in
 non-minimal UED scenarios (for the case of the $A_3^{(2)}$ resonance, we show instead
 the dominant decay to SM particles if the decay into KK(1) leptons was {\it not} kinematically 
 allowed).
The other important parameters to take into account are clearly the (effective) couplings that 
appear in Fig.~\ref{fig:feynbbtogammax}. In Tab.~\ref{tab:resonances}, we thus also indicate for 
reference the size of the (effective) $B^{(1)}B^{(1)}Y^{(2)}$ coupling as well as the decay rate 
$Y^{(2)}\to\gamma X$. From this overview, it becomes clear that the $H^{(2)}$ resonance is 
clearly expected to result in the strongest line signal: it is not only the most long-lived 
resonance, but also the only one that couples to the incoming LKP pair at tree level.

\begin{figure}[t]
\begin{center}
 \includegraphics[width=\columnwidth]{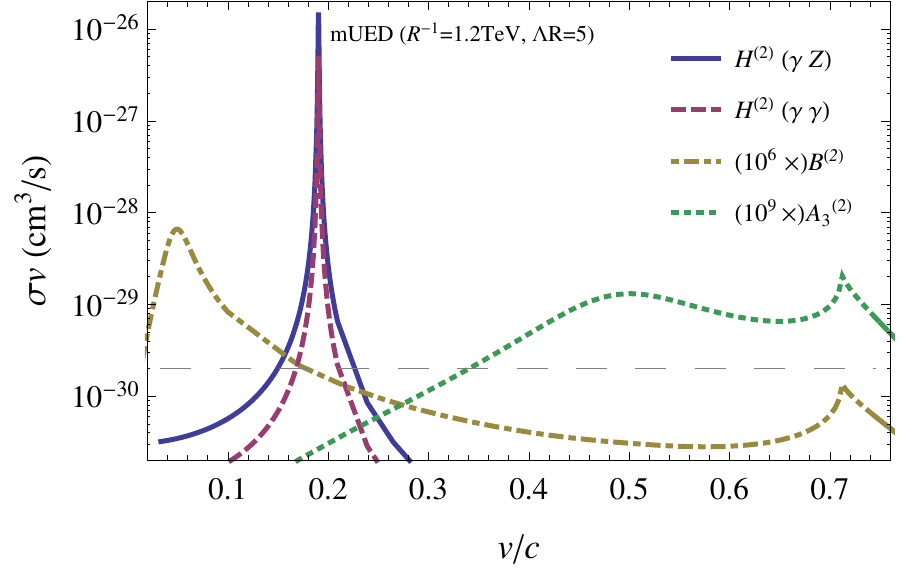}
\end{center}
\caption{\label{fig:svall}$B^{(1)}B^{(1)}\to\gamma X$  cross sections in the mUED scenario, for 
the various channels considered in Fig.~\ref{fig:feynbbtogammax}, as a function of the relative 
speed of the WIMPs (the curves associated with the $B^{(2)}$ and 
$A_3^{(2)}$ resonances are multiplied by factors $10^6$ and $10^9$ respectively). The 
horizontal 
dashed line indicates the dominant line signal in the zero-velocity limit, which arises 
from $B^{(1)}B^{(1)}\to\gamma\gamma$ \cite{Bergstrom:2004nr}. Note that the {\it location} of 
the resonances is essentially a free parameter in UED theories; in particular, it can occur at much 
smaller velocities than shown here for the mUED case.} 
\end{figure}

We have performed a full calculation of the dominant contribution to all annihilation processes 
shown in Fig.~\ref{fig:feynbbtogammax}, which includes a determination of the relevant effective 
couplings and decay rates (for details, see Appendices \ref{app:KKdec} and \ref{app:KKann}). In 
Fig.~\ref{fig:svall}, we show the individual contributions to the cross section for 
$B^{(1)}B^{(1)}\rightarrow \gamma X$ from these diagrams. Note that the ratios of the peak
values agree well, within an order of magnitude, 
with the naive estimates one can infer directly from the values stated in 
Tab.~\ref{tab:resonances}. In particular, the by far largest 
cross section for a monochromatic photon can be obtained for 
$B^{(1)}B^{(1)}\stackrel{H^{(2)}}{\longrightarrow} \gamma Z$, with a very pronounced resonance 
corresponding to the mass of the ${H^{(2)}}$. Remarkably, this cross section (as well as the 
corresponding process for $\gamma\gamma$ final states) can be significantly larger than the 
cross section for $B^{(1)}B^{(1)}\rightarrow \gamma X$ in the zero velocity limit as  indicated by 
the dashed line, $\sigma v=2 \times 10^{-30}\, \rm cm^3/s$. In fact, even at $v=0$, the 
$H^{(2)}$ resonance thus contributes at roughly the same level as $\gamma Z$ final states 
without taking into account these contributions \cite{Bergstrom:2004nr,Bertone:2010fn}.
While the {\it locations} of the 
resonances are specific to the mUED scenario, the 
couplings are typically only affected at the level of radiative corrections for deviations from the 
minimal scenario; this implies that the {\it signal strengths} shown in this figure are rather generic.
% changing coupling in non-mUED: possible but not intrinsically related to mass splittings
% (only change counter terms for self-energy {\it on the brane}) $\leadsto$ Small corrections for 
% couplings that do not violate KK number; otherwise in principle arbitrary. But: increasing 
% KK-number violating couplings would both increase the decay rate to charged SM particles 
% (denominator) {\it and} to $\gamma X$ (nominator). So additional fine-tuning needed to
% significantly alter results.
A possible exception to this last comment would occur if the mass spectrum displayed in 
Fig.~\ref{fig:atomicphysplot} would change in a qualitative way, opening up new or closing 
existing 
decay channels. An interesting possibility to even further enhance the 
$B^{(1)}B^{(1)}\rightarrow \gamma X$ rate beyond the mUED expectation would also be to 
increase the mixing between the KK top quark states beyond its mUED value of 
$\sin2\alpha_t^{(1)}=0.143$, a quantity which enters quadratically in the cross section
 (\ref{Rhgg}, \ref{RhgZ}).

\begin{figure}[t!]
\begin{center}
 \includegraphics[width=\columnwidth]{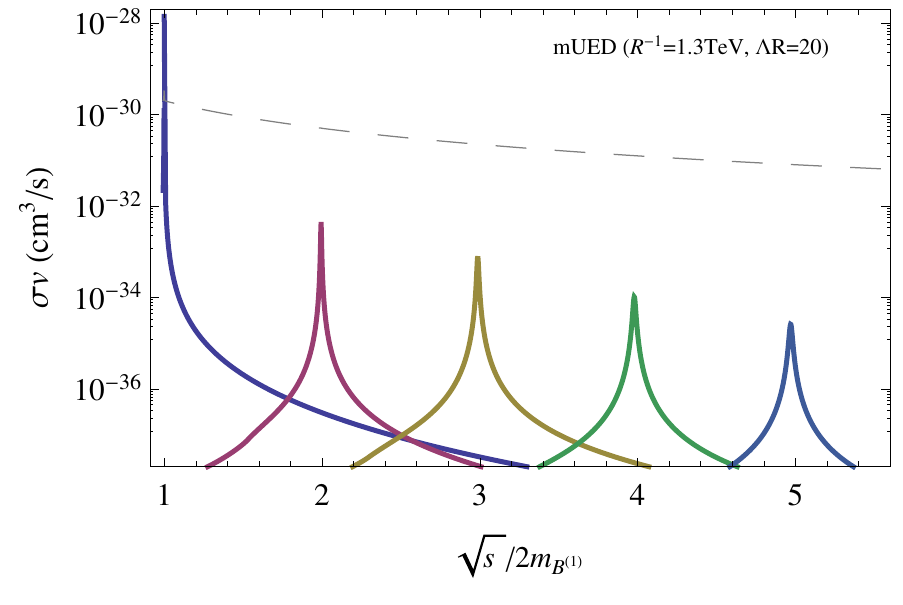}
\end{center}
\caption{\label{fig:higherres} Contributions of the first five $H^{(2n)}$ resonances to the 
$B^{(1)}B^{(1)}\to\gamma Z$ annihilation rate in the mUED scenario, taking however $\Lambda R=20$, 
as a function of $\sqrt{s}/2m_{B^{(1)}}$. The dashed line is an extrapolation 
($\propto s^{-1}$) of the standard result for $B^{(1)}B^{(1)}\to\gamma\gamma$~\cite{Bergstrom:2004nr}. 
Note that in general one can encounter much larger peak normalizations than what is shown here for the 
{\it  minimal} UED case. 
%dominated by H^2n, B^4 and A^4 are strongly suppressed by sin\alpha^(4)
} 
\end{figure}

So far, we have only mentioned the effect of second-level KK resonances. As discussed in 
Section  \ref{sec:he} later on, however, there may exist extreme astrophysical environments 
where much higher CMS energies  are available for the collision of two LKPs. If those energies 
are sufficient to excite higher KK resonances $Y^{(2n)}$, with $n>1$, this would lead to a rich 
phenomenology. While we do not aim at an exhaustive discussion here, we would like to point 
out that most of the 
arguments presented above can straightforwardly be applied to this situation as well. One of the 
most striking consequences, however, may in any case be the appearance of {\it multiple 
gamma ray lines with an equidistant spacing in energy} that equals almost exactly twice the 
inverse of the compactification radius, $\Delta E_\gamma^\text{line}\simeq 2/R$: if such a 
striking spectral signature would be observed, this would constitute a smoking gun signal for the 
higher-dimensional origin of the cosmological DM. 

In Fig.~\ref{fig:higherres}, we show explicitly  that such a structure indeed appears in the mUED model.\footnote{
Note that in order to demonstrate this effect, we have allowed for a larger cutoff value  
$\Lambda R=20$, which can be motivated by slightly non-minimal boundary terms affecting the 
Higgs self-coupling and thus circumventing the arguments from vacuum stability that lead to 
$\Lambda R\sim5$.} 
While  $H^{(2n)}$ resonances dominate over other resonances also at $n>1$, however, their  
contribution  to the annihilation cross-section is a few orders of magnitude smaller than that of 
the first resonance (shown in more detail in Fig.~\ref{fig:svall}).
This is mainly due to two reasons. First, in contrast to the tree-level coupling between $H^{(2)}$ 
 and the two incoming LKPs, the corresponding couplings of higher-level  $H^{(2n)}$ are 
radiatively generated and localized at the orbifold fixed points -- see Appendix 
\ref{app:effvert} for an exhaustive discussion on radiative vertices in UED. Secondly, higher-level 
KK states have more possible decay modes and it is thus less likely that there is no 
kinematically allowed channel at tree-level. As discussed above, the resulting larger decay 
widths thus decrease the expected peak annihilation rate. Relaxing the restrictive assumptions 
of the mUED model, however, both these caveats can be overcome: allowing large boundary 
terms at the cut-off scale that prevent the $B^{(1)}B^{(1)}H^{(2n)}$ ($n>1$) vertex from being 
strictly radiative, while keeping the vertices relevant for the $H^{(2n)}$ decay small,  would boost 
 these resonances to a level which can be fairly comparable to the one encountered in the 
 $H^{(2)}$ case.

To summarize this Section, we have pointed out the remarkable fact that resonances in 
extra-dimensional DM models may naturally enhance monochromatic gamma ray lines much 
more than the  continuum signal of secondary and FSR photons. In the mUED model, this 
leads to a line signal which for DM velocities $v\sim\mathcal{O}(0.1)$ can be enhanced by 
several orders of magnitude with respect to the $v=0$ limit known from the literature. 
As we have stressed, however, there is no particularly strong reason to adopt the restrictive 
limitations of the minimal model. In fact, UED scenarios generally allow for resonances at 
considerably smaller relative velocities of the annihilating DM particles and, to a lesser degree, 
there is also some freedom to enhance the  signal normalization even further. Potentially even 
more important, these scenarios allow for the appearance of multiple strong line signals at 
equally spaced photon energies.

%%%%%%%%%%%%%%%%%%%%%%%%%%%%%%%%%%%%%%%%%%%%%
\section{Gamma ray flux from DM annihilation}
\label{sec:astro}

Neglecting for the moment relativistic effects, the expected gamma ray flux from DM annihilation 
from a direction $\psi$, averaged over the opening angle $\Delta \psi$ of the detector, is given 
by:
\bea
\displaystyle\frac{\der\Phi}{\der E_{\gamma}}(E_\gamma,\psi)&=& \frac{1}{8 \pi} \int_{\psi} 
\frac{\der \Omega}{\Delta \psi}\int_\text{l.o.s} \der l(\psi) \rho^2(\bvec r)\times \nonumber\\
&&\displaystyle \times \frac{1}{m^2_{\rm DM}}\left\langle\sigma v_{\rm rel} \sum_f B_f
\frac{\der N^f_\gamma}{\der E_\gamma}\right\rangle\,,
\label{eq:flux}
\eea
where the integration is performed along the line of sight (l.o.s) and we take into account that 
the DM particles are self-conjugate (for DM candidates with distinct particle and anti-particle 
there would be an additional factor of 1/2). The spatial distribution of the signal traces the DM 
density profile $\rho(r)$ and is typically assumed to be fully determined by the quantity 
$J(\psi) \equiv \int_{\psi} \der \Omega/\Delta \psi\int_{l.o.s} \der l(\psi) \rho^2(r)$. The second line 
of Eq.~(\ref{eq:flux}) contains the particle physics of the underlying theoretical model, as 
discussed in the previous Section: the velocity-weighted CMS annihilation cross section 
$\sigma v_{\rm rel}$ at present time, the branching ratio $B_f$ into channel $f$, times the 
number of photons $N^f_\gamma$ produced per annihilation; this factor determines the spectral 
shape of the signal.

The velocity-average that appears above is given by
\bea
 \langle f(v_{\rm rel})\rangle &\equiv&
  \int\!\!\der{}^3v_1 \der{}^3v_2P_{ \bvec r}(\bvec v_1)P_{ \bvec r}(\bvec v_2) f(v_{\rm rel})
  \nonumber\\
  &=&
  \int\!\!\der{}^3v_{\rm rel} P_{\bvec{r},\rm{rel}}(\bvec v_{\rm rel})f(v_{\rm rel})\\
  &\equiv& \int\!\!\der v_{\rm rel} \, p_{\bvec{r},\rm{rel}}( v_{\rm rel})f(v_{\rm rel})\,,
\eea
where $P_{\bvec r}(\bvec v_i)$ is the $3$D normalized velocity distribution function of a WIMP 
at a position $\bvec r$ and 
\begin{equation}\label{eq:vrel}
P_{\bvec{r},\rm{rel}}(\bvec v_{\rm rel})\equiv\int\!\!\der{}^3v_{\rm CM} 
P_{ \bvec r}(\bvec v_{\rm CM}+\bvec v_{\rm rel}/2)P_{\bvec r}(\bvec v_{\rm CM}-\bvec v_{\rm rel}/2) 
\end{equation}
 is the $3$D distribution function of the relative velocities of the WIMPs, with 
 $\bvec v_{\rm rel}\equiv\bvec v_1-\bvec v_2$ and $\bvec v_{\rm CM}\equiv(\bvec v_1+
 \bvec v_2)/2$. 
 For a Maxwell-Boltzmann (MB) distribution $p_{\bvec r}(v)= 4\pi^{-\frac12}v_0^{-3}v^2\exp[-v^2/v_0^2]$  
 with most probable velocity 
 $v_0$, for example,  $p_{\bvec{r},\rm{rel}}(\bvec v_{\rm rel})$ is given by an MB distribution 
 with most probable velocity $\sqrt{2}v_0$.
 
 Let us stress that  in general the particle physics factor  in Eq.~(\ref{eq:flux}) cannot be 
 factorized out of the integral along the line of sight, because an implicit $\bvec{r}$-dependence 
 enters via the velocity-average over $P_{\bvec{r},\rm{rel}}(\bvec v_{\rm rel})$. In our case, the 
 gamma ray flux thus depends on both the density and velocity distribution profile of the DM 
 particles -- in contrast to the typically assumed situation of a velocity-independent cross section. 
 
The general expression (\ref{eq:flux}) simplifies for the monochromatic photons emitted by the 
annihilation of LKP pairs we consider here. The differential cross section near a  strong 
resonance, in particular, is given by
\beq
  \sigma\sum_f B_f\frac{\der N^f_\gamma}{\der E_\gamma}\simeq
  \frac{N_\gamma\sigma_\text{res}\Gamma_\text{res}^2m_\text{res}^2}{(s-m_\text{res}^2)^2+
  \Gamma_\text{res}^2m_\text{res}^2}\delta\left(E_\gamma-\frac{\sqrt{s}}{2}\right)\,,
\eeq
where $m_\text{res}$ and $\Gamma_\text{res}$ are the mass and width of the $s$-channel 
particle, respectively, $\sigma_\text{res}$ is the peak value of the cross section and we have 
neglected the masses of the final state particles; $N_\gamma=2$ for $\gamma\gamma$ final 
states and $N_\gamma=1$ otherwise. Given that $s=4m_\text{DM}^2/(1-v_\text{rel}^2/4)$ in the 
CMS system,
%\beq
%\frac{s}{4m_\text{DM}^2}=1+\frac{1}{4}v_\text{rel}^2+\mathcal{O}(v^4)\,,
%\eeq
%% NB: there is no contribution from $v_CM$ at quadratic order !!!
the velocity-average can  be evaluated trivially and the flux near the resonance becomes
\beq
\displaystyle\frac{\der\Phi_\text{res}}{\der E_{\gamma}}\simeq\frac{N_\gamma
\left(\sigma v_{\rm rel}\right)_\text{res}}{8\pi E_\gamma^3}
\frac{\tilde J_\text{line}(E_\gamma)\,\Gamma_\text{res}^2m_\text{res}^2}{(4E_\gamma^2 -
m_\text{res}^2)^2+\Gamma_\text{res}^2m_\text{res}^2}\,,
\label{eq:phires}
\eeq
where
\beq
\tilde J_{\rm line}(E_\gamma)\equiv \frac1{\Delta\Omega}\int_{\Delta\Omega}\!\!\!\!\der\Omega
\int_{\rm l.o.s.}\!\!\!\!\!\der l\int\rho^2(\bvec r)\frac{p_{\bvec{r},\text{rel}}(v_\text{rel}^\text{line})}
{v_\text{rel}^\text{line}/4}
\eeq
and $v_\text{rel}^\text{line}(E_\gamma)=2\sqrt{1-m_\text{DM}^2/E_\gamma^2}\simeq
\sqrt{8}\sqrt{E_\gamma/m_\text{DM}-1}$.

Before discussing in more detail the implications for the UED model, however, 
let us in the following subsections briefly describe the DM density profile and the DM velocity 
distribution in a halo that enter in the above expressions.

%%%%%%%%%%%%%%%%%%%%%%%%%%%%%%%%%%%%%%%%%%%%%
\subsection{DM density profile}\label{sec:rho}

A generic parametrization for a spherically symmetric DM density, that encompasses several 
halo profiles, is given by:
\begin{equation}
\rho(r) = \rho_\odot \left(\frac{r}{r_\odot}\right)^{-\gamma} \left[\frac{1+(r_\odot/r_0)^\alpha}{1+(r/r_0)^\alpha}\right]^{\frac{\beta-\gamma}{\alpha}}\,,
\end{equation} 
where $\rho_\odot \simeq (0.3-0.4)\, {\rm GeV/cm^3}$~\cite{Salucci:2010qr,Bovy:2012tw} is the 
DM density in the solar neighborhoods and $r_\odot \simeq 8.5$ kpc denotes the Sun's position 
with respect to the GC. In our analysis we consider the two density profiles with the extreme 
opposite behavior in the inner part. The NFW density profile \cite{Navarro:1995iw} is obtained 
for $\left(\alpha, \beta, \gamma\right)  = (1,3,1)$ and a scale radius of $r_0 = 20$ kpc. Such a 
cuspy profile is favored by numerical $N$-body simulations; for very small galactocentric 
distances, $r\lesssim100$\,pc, the profile may in fact be even steeper and exhibit a slope of 
$\rho\propto r^{-1.2}$ \cite{Diemand:2008in,Diemand:2005wv,Stadel:2008pn}.
 On the contrary the cored isothermal profile -- which is observationally inferred for low surface 
 brightness as well as dwarf galaxies \cite{deNaray:2011hy,Walker:2011zu} -- has 
 $\left(\alpha,\beta,\gamma\right) = \left(1,2,1\right)$, a scale radius of 3.5 kpc and  a finite 
 density core close to the GC. 

It is very likely that the central SMBHs in galaxies have primordial DM density 
spikes~\cite{Gondolo:1999ef}. When indeed a BH forms, the DM distribution adjusts to the new 
gravitational potential and this process leads to the formation of spikes. Even though 
gravitational scattering off stars of DM particles and DM 
annihilation~\cite{Bertone:2005xv,Bertone:2005hw} 
tend to reduce the DM density in spikes, the enhancement is still significant with respect to 
ordinary cuspy profiles. This leads to a change in the slope $\gamma\to\gamma'=7/3$ of the  
density profile within the BH 
radius of influence for a NFW profile (prior  to  BH formation), with a plateau at a radius where 
annihilations become important over the BH life-time. GR corrections to the profile for the 
Schwarzschild case are computed in~\cite{Sadeghian:2013laa}, where the inner radius of the 
annihilation plateau is found to be $2 r_S$, where $r_S$ is the Schwarzschild radius of the BH. 
The extent to which the spikes  survive dynamical heating by their environment is unknown. We 
consider the pessimistic case described in~\cite{Gnedin:2003rj}, where an initially  
$\gamma'=3/2$ profile 
is adopted, arising from a cored isothermal profile.  In other words, in the case of the GC, stellar 
scattering affects the BH spike over several  core relaxation times, amounting to a few  Gyr. 
Hence the density profile is most likely softened to a $\gamma'\sim 3/2$ 
profile~\cite{Vasiliev:2008uz}. 
However more massive SMBH dynamical relaxation  via stellar interactions does not occur, 
because the core relation time-scales are much longer. In the case of M87, which we will 
discuss later as one of our most promising candidates, the core-related time-scale is of order 
$10^5$ Gyr. Hence the initial spike profile is preserved.

More quantitatively, consider a DM density spike surrounding a massive BH. There are several 
scales of interest. The gravitational radius of influence, which by definition contains the 
same mass as the BH, and the half-mass 
radius of the spheroid are respectively:
\begin{eqnarray}
 r_i= \frac{GM_{\rm BH}}{\sigma^2}\nonumber\,, \\
 r_{1/2}=\frac{GM_{1/2}}{\sigma^2}\,,
 \end{eqnarray}
where $\sigma$ is the bulge velocity dispersion. The density profile is then given by
\begin{eqnarray}
\rho \propto
\left\{
\begin{array}{rl}
r^{-\gamma}  & \mbox{if } r>r_i\,, \\
 r^{-\gamma'} & \mbox{if } r<r_i\,.
\end{array}
\right.
\end{eqnarray}
Efficient annihilation sets an upper limit on the DM density in the innermost parts. 
A density plateau, with $\rho_p\equiv\rho(r\lesssim r_p)$, thus occurs at a radius $r_p$ where the 
annihilation timescale equals the BH age, $t_{\rm BH} \sim 10^8-10^{10}$ years, i.e. where 
\begin{eqnarray}\label{eq:plateau}
\rho(r_p)\, (\sigma v_{\rm rel}) &  = &  \frac{m_{\rm DM}}{ t_{\rm BH}}\,.
\end{eqnarray}

%%%%%%%%%%%%%%%%%%%%%%%%%%%%%%%%%%%%%%%%%%%%%
\subsection{Velocity distribution}\label{sec:vel}

By definition the density profile and the velocity distribution are related via
\begin{equation}
\rho (\bvec{r}) = m_{\rm DM} \int {\rm d}^3 v \  F(\bvec{v},\bvec{r})\, ,
\label{eq:Fv1}
\end{equation}
where $F(\bvec{v},\bvec{r})$ is the WIMP phase space distribution in the Galactic frame. Given 
a DM density profile, the underlying DM velocity distribution 
% $P_{\bvec r}(\bvec v_i)= \int d^3r'\, F(\bvec{v},\bvec{r'}) / (\rho (\bvec{r})/m_{\rm DM})$ 
can be extracted by inverting 
Eq.~(\ref{eq:Fv1})  under the assumption of hydrostatic equilibrium, a solution known as the 
Eddington formula~\cite{Binneybook}. The above integral can be inverted only under certain 
assumptions, such as spherical symmetry for the density profiles. For instance the MB 
distribution results from an isothermal density profile  scaling as $r^{-2}$.

Other spherically symmetry density profiles, such as NFW, can be used to infer the 
corresponding velocity distribution and an application of this procedure for DM  indirect detection 
is given in~\cite{Ferrer:2013cla}. However it has been shown in~\cite{Ling:2009eh} that the 
presence of  baryons in N-body simulations has the effect of making the matter distribution more 
concentrated by adiabatic contraction, and the WIMP velocity distribution is brought closer to a 
MB distribution. Significant departure from the MB on the other hand arises when the velocity 
dispersion becomes small, e.g.~close to the GC or in dwarf galaxies. In these regions, however, 
the main uncertainty derives from the inner slope of the density profile,  which is difficult to 
extract from data and has not  converged in  simulations either. In fact, these $N$-body 
simulations indicate that DM halos are anisotropic and exhibit clumpy structures and 
streams, features that cannot be  captured by the Eddigton formula in a simple way. 
Going even closer to the central BH, where the DM spikes form, the assumption of hydrostatic 
equilibrium is not satisfied anymore: in this case, one would have to extract the DM velocity 
distribution from $N$-body simulations after adiabatic contraction. 
In order to avoid addressing in detail the 
large uncertainties involved in any brute force computation of the velocity distribution near the 
SMBH, we  will in the following make the simplifying assumption of a MB distribution when 
considering DM particle collisions.

The old star population  ($> 1$ Gyr) in the central 0.5 pc  of our galaxy has a stellar cusp with 
relatively shallow  slope $n(r) \propto r^{-\gamma}, $  where 
$\gamma= 0.4 \pm 0.2$~\cite{2013ApJ...779L...6D}, measured in a three-dimensional kinematic 
study. This slope is much flatter than the dynamically relaxed expectation 
($\gamma = 3/2 - 7/4$)  that we have adopted for the DM. The flattening  is attributed to stellar 
heating. We note however that the recently discovered~\cite{Schodel:2014gna} 
young nuclear star cluster centered on SagA* ($\simgt 50\%$ of the stars formed in the most 
recent star formation event  $2-6$ Myr ago) 
has a significantly steeper slope within its half-light radius of $\sim 4 $pc, comparable to the 
gravitational sphere of influence radius at $\sim 3$ pc. Because the BH has certainly grown by 
accretion of gas and 
stars over the past Gyr, it  is not clear how the competing effects of adiabatic contraction of the 
DM, that steepens the profile, compete with dynamical heating. As discussed previously, the 
effects of stellar heating are irrelevant for SMBHs much more massive than in our galactic  
center, as is the case for M87 and Cen A. The three-dimensional kinematic study shows that the 
velocity field is consistent with that earlier inferred from orbital studies within 0.05 pc of SagA* 
and yields a similar mass estimate for the central SMBH. Our simplifying assumption of a MB 
distribution should not modify our estimates of collision velocities by a significant factor 
compared to the other uncertainties in our model.

%%%%%%%%%%%%%%%%%%%%%%%%%%%%%%%%%%%%%%%%%%%%%
\section{Enhanced gamma ray lines from KK DM annihilation}
\label{sec:le}

\begin{figure}[t!]
\begin{center}
 \includegraphics[height=0.8\columnwidth]{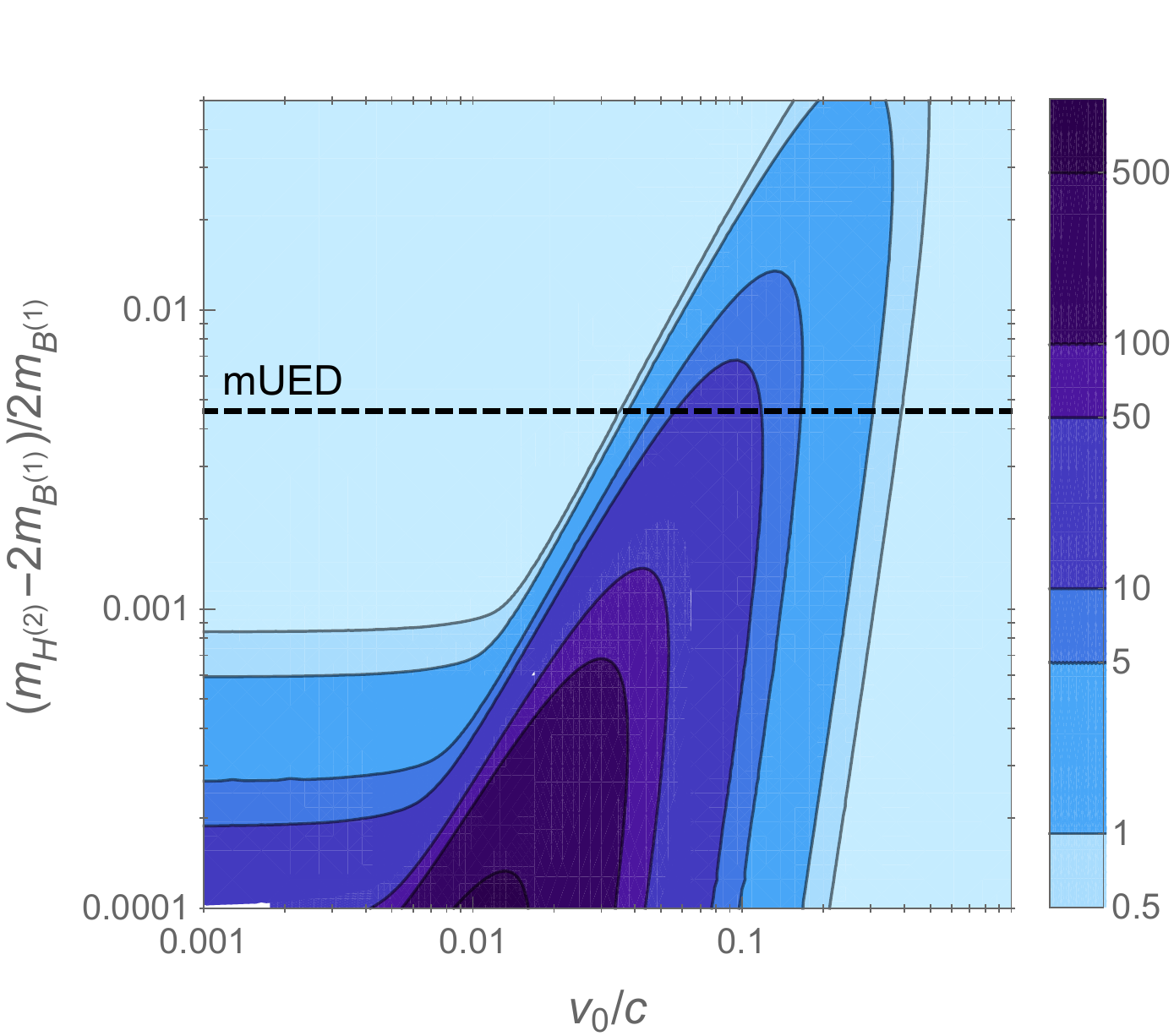}
\end{center}
\caption{\label{fig:dmv} Enhancement of the line signal due to the presence of resonances: the 
color scale indicates the ratio of the monochromatic photon flux resulting from the  diagrams 
shown in Fig.~\ref{fig:feynbbtogammax}, assuming a Mawell-Boltzmann distribution with most 
probable DM velocity $v_0$, to the one expected in the $v=0$ limit. The dashed line indicates the 
mass difference between the resonance $H^{(2)}$ and twice the LKP $B^{(1)}$ in the minimal 
scenario (for $R^{-1}=1.2$\,TeV and $\Lambda R=5$).} 
%mUED: sigmagg0=1.99*10^-30
\end{figure}

Let us now assess in more detail by how much the line signal at $E_\gamma\sim m_{B^{(1)}}$ 
can be enhanced with respect to the $v=0$ result in concrete applications. To this end, we 
assume a MB distribution for the WIMP velocities and compare the flux from resonant 
diagrams, as given by Eq.~(\ref{eq:phires}), with the flux in the zero velocity limit. The result is 
shown in Fig.~\ref{fig:dmv} as a function of the resonance mass and the assumed velocity 
dispersion
(using for example the velocity dispersion extracted from 
Fig.~18 of~\cite{Ferrer:2013cla} instead of MB, based on an NFW profile with a baryonic bulge, 
does not qualitatively change our results).
For an $H^{(2)}$ mass as given in the mUED scenario, indicated by the dashed line, the hitherto 
neglected diagrams will thus enhance the line signal by more than an order of magnitude for a 
most probable % individual
DM velocity of $v_0\gtrsim0.05$. Allowing for a smaller $H^{(2)}$ mass in non-minimal 
scenarios, on the other hand, will result in similar enhancements already for much smaller 
velocities. In the most favourable case, the signal may be up to three orders of magnitude 
stronger than expected from the calculation in the $v=0$ limit.

It should be noted that the typical velocities in the {\it Milky Way} are of the order of
$v\sim10^{-3}$, such that there is seemingly little hope to actually observe resonant LKP 
annihilation on galactic scales (see, however, the next Section). Furthermore, current limits on 
monochromatic photons at TeV energies \cite{Abramowski:2013ax}, deriving from observations 
of the Galactic center, are almost three orders of magnitude weaker than the mUED signal 
expected in the $v=0$ limit (though CTA will improve these limits by significantly more than one 
order of magnitude~\cite{CTAurl}). While $s$-channel annihilation will in general start to be 
important for mass differences between the $H^{(2)}$ mass and two times the LKP mass at the 
per mille level, as illustrated in Fig.~\ref{fig:dmv},  even more degenerate spectra would thus be 
needed to make the signal observable with near future technology. 

The situation is considerably better for {\it galaxy clusters}, where the typical velocities of up to  
$v\sim10^{-2}$ would be sufficient to probe the resonant regime for less non-minimal 
setups.  Another advantage of clusters is that they are the astronomical targets for indirect DM 
searches with the largest mass hierarchy of (sub)sub halos, which implies that they maximise 
the signal enhancement due to substructures \cite{Pinzke:2011ek}. In optimistic scenarios for 
the distribution of substructures, they may thus be the brightest sources of DM annihilation 
radiation \cite{SanchezConde:2011ap}. In combination with the enhancement from resonances 
studied here, line signals from clusters may thus offer a promising opportunity for DM detection 
with future Air Cherenkov telescopes like CTA.

Still, even the velocity dispersion of  $0.008c$ measured in the most massive galaxy cluster 
known~\cite{2008A&A...481..593M}  appears relatively small in our context, and to really probe 
the extra-dimensional resonances even higher particle velocities are in general indispensable.
Hence we turn next to an environment where there are known sources of high energy 
gamma rays and high particle collision velocities are inevitable, in the vicinity of SMBHs.

%%%%%%%%%%%%%%%%%%%%%%%%%%%%%%%%%%%%%%%%%%%%%
\section{Gamma rays from dark matter spikes: Schwarzschild black hole case}
\label{sec:he}

SMBHs are found to be effective DM particle accelerators. Particle collisions occur to high CMS 
energies and are especially important for the case of a Kerr BH~\cite{Banados:2009pr}. The 
CMS energy for particle collisions is limited to 4.5 times the rest mass  for a Schwarzschild  BH, 
in the case of Kerr attains energies of  20 times the rest mass for Kerr parameter\footnote{$a$ is 
the angular momentum $J$ divided by the BH mass.} $a=0.998$, and is formally  infinite for the 
extremal Kerr BH.

The following issues have been raised with regard to whether there is any potentially observable 
flux, namely whether the density spike  survives, whether there is a negligibly small flux at  
infinity, whether back-reaction limits the acceleration, and whether the large red-shifting of 
photons generated in DM particle collisions renders any debris 
unobservable~\cite{Jacobson:2009zg,Berti:2009bk,McWilliams:2012nx,2012PhRvL.109l1101B}.

None of these issues are insuperable for a number of reasons, none of which can however be 
considered definitive, but are discussed in~\cite{Vasiliev:2008uz,2012JHEP...12..032Z,Grib:2010xj,2013PhRvD..88f4036T,Zaslavskii:2013et}. 

At the very least, one can infer that the topic  of particle collision 
signatures near BH horizons merits further study. Specifically, several ways have been proposed for 
observing BH-boosted annihilations:

\begin{enumerate}
\item  It has been shown that there are some unbound  geodesics around the axis of 
rotation~\cite{2013ApJ...774..109G}. If the ergosphere were evenly populated with injection of 
annihilation debris, an increasing fraction of null geodesics are unbound in the limit of large and 
increasing $a$. To feed these would require that annihilation seeds  the Penrose effect.
\item Penrose boosting of the energetics of collisional debris can occur in the ergosphere. 
Sufficiently detailed models have not been worked out however to come to a quantitative 
result~\cite{2012PhRvD..86b4027H, 2012PhRvL.109l1101B,Zaslavskii:2012yp}.
\item There is no horizon near naked singularities. Dirty BH are another option. In these cases 
collisions at infinite CMS energies are possible~\cite{Zaslavskii:2013nra}.
\end{enumerate}
Given the considerable interest should any signal be observable, and that the possibility of such 
an effect remains to be clarified, we have decided to explore a potentially unique signal from KK 
particle annihilations near the horizon. For our further considerations we stick for simplicity to the 
case of a non-rotating Schwarzschild BH. Because of the spherical symmetry of the system, it is 
enough to study the collision of DM particles in the equatorial plane to recover the general 
solution.

Considering the DM density around a BH, described by the spike and plateau configuration, a 
key scale ratio is that of plateau scale to Schwarzschild radius,
\begin{equation}
\frac{r_p}{r_S}=\left(c\over\sigma\right)^2 \left({{\langle\sigma v_{\rm rel}\rangle t}\over m_{\rm DM}}\rho_{1/2}\right)^{1/\gamma'}\left(M_{1/2}\over M_{\rm BH}\right)^{\gamma/\gamma'}\,.
 \end{equation}
The plateau radius approaches the horizon for the most massive BHs: this amplifies the 
annihilation flux considerably.

The maximum luminosity (number of $\gamma$s per second) is evaluated at $r_h=2r_S$,
\begin{equation}
L_h={4\pi\over 3}{(2r_S)^3\over \langle\sigma v_{\rm rel}\rangle t^2}\,,
\end{equation}
and the total spike luminosity is
\begin{equation}
L_{sp} =L_h\left({r_p\over 2 r_S}\right)^3 \,.
\end{equation}
This reduces to
\begin{eqnarray} 
L_{sp} & = &
\left(c \over \sigma\right)^6 \left({{\langle \sigma v_{\rm rel} \rangle t}\over m_{\rm DM}}\rho_{1/2}\right)^{3/\gamma'}
\left(M_{1/2}\over M_{\rm BH}\right)^{3\gamma/\gamma'}\nonumber\\
& & \times { 32\pi\over 3}{r_S^3\over{\langle \sigma v_{\rm rel} \rangle t^2}}\,.
\end{eqnarray}
For scaling purposes, we assume $\sigma^4=G^2\Sigma_{1/2} M_{1/2}$ 
and $\rho_{1/2}=\Sigma_{1/2}^{3/2}M_{1/2}^{-1/2}$, where $\Sigma_{1/2}=\rho_{1/2}r_{1/2}$. 
We assume $\sigma$ is unchanged between $r_{sp}$ and $r_{1/2}$. We now find 
  \bea
  L_{sp} ={ 32\pi\over 3}{{\langle \sigma v_{\rm rel} \rangle}^{3/\gamma'-1}\over m_{\rm DM}^{\gamma/\gamma'}}t^{(3/\gamma'-2)}
  \Sigma_{1/2}^{{3\over {2}}(3/\gamma'-1)}\cr
 \times  \left(M_{BH}\over M_{1/2}\right )^{3(1-\gamma/\gamma')}M_{1/2}^{{3\over 2}(1-1/\gamma')}.
\eea 
Let us consider the case of a NFW halo profile: in the central region, $\gamma=1$ and $\gamma'=7/3$. Here we have 
\begin{equation}
 L_{sp} ={ 32\pi\over 3}{{\langle \sigma v_{\rm rel} \rangle}^{2/7}\over m_{\rm DM}^{9/7}}t^{-5/7}
  \Sigma_{1/2}^{3/7}  \left(M_{\rm BH}\over M_{1/2}\right )^{6/7}M_{\rm BH}^{6/7}\,.
\end{equation}
One can now see the explicit dependence on BH mass $M_{\rm BH}$. The dependence on 
$M_{BH}/M_{1/2}$ is found to be constant at the present epoch (and to be reduced weakly with 
increasing redshift)~\cite{2013arXiv1304.7762K}. The empirical dependence of surface 
brightness on galaxy luminosity,  both defined at the effective radius (equivalent to a correlation 
between $\Sigma_{1/2}$ and $M_{1/2}$) is also weak~\cite{2012ApJS..198....2K}.

\begin{figure*}[t]
\begin{minipage}[t]{0.44\textwidth}
\centering
\includegraphics[width=1.\columnwidth,trim= 20mm 3mm 20mm 13mm, clip]{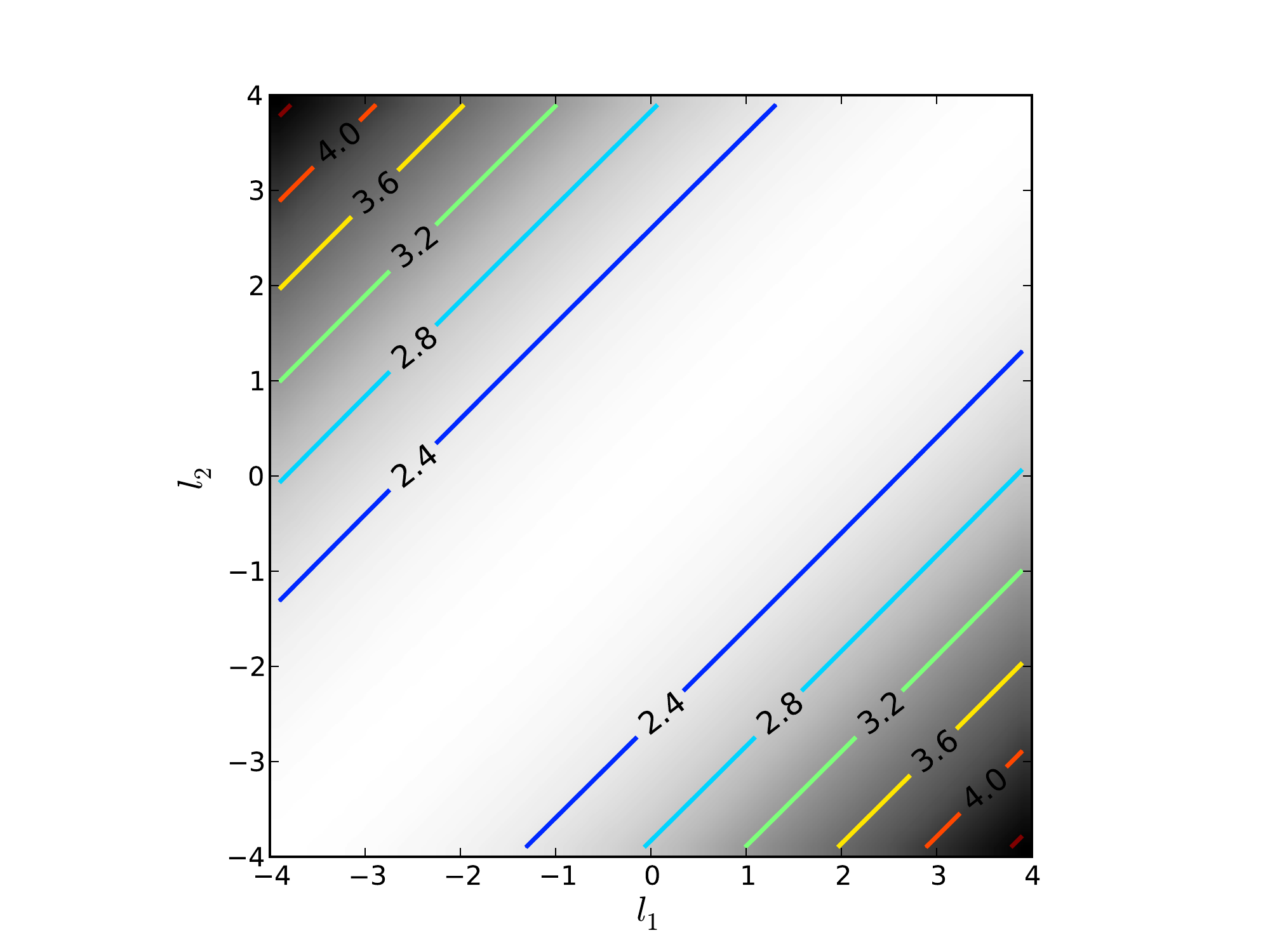}
\end{minipage}
\begin{minipage}[t]{0.44\textwidth}
\centering
\includegraphics[width=1.\columnwidth,trim= 20mm 3mm 20mm 13mm, clip]{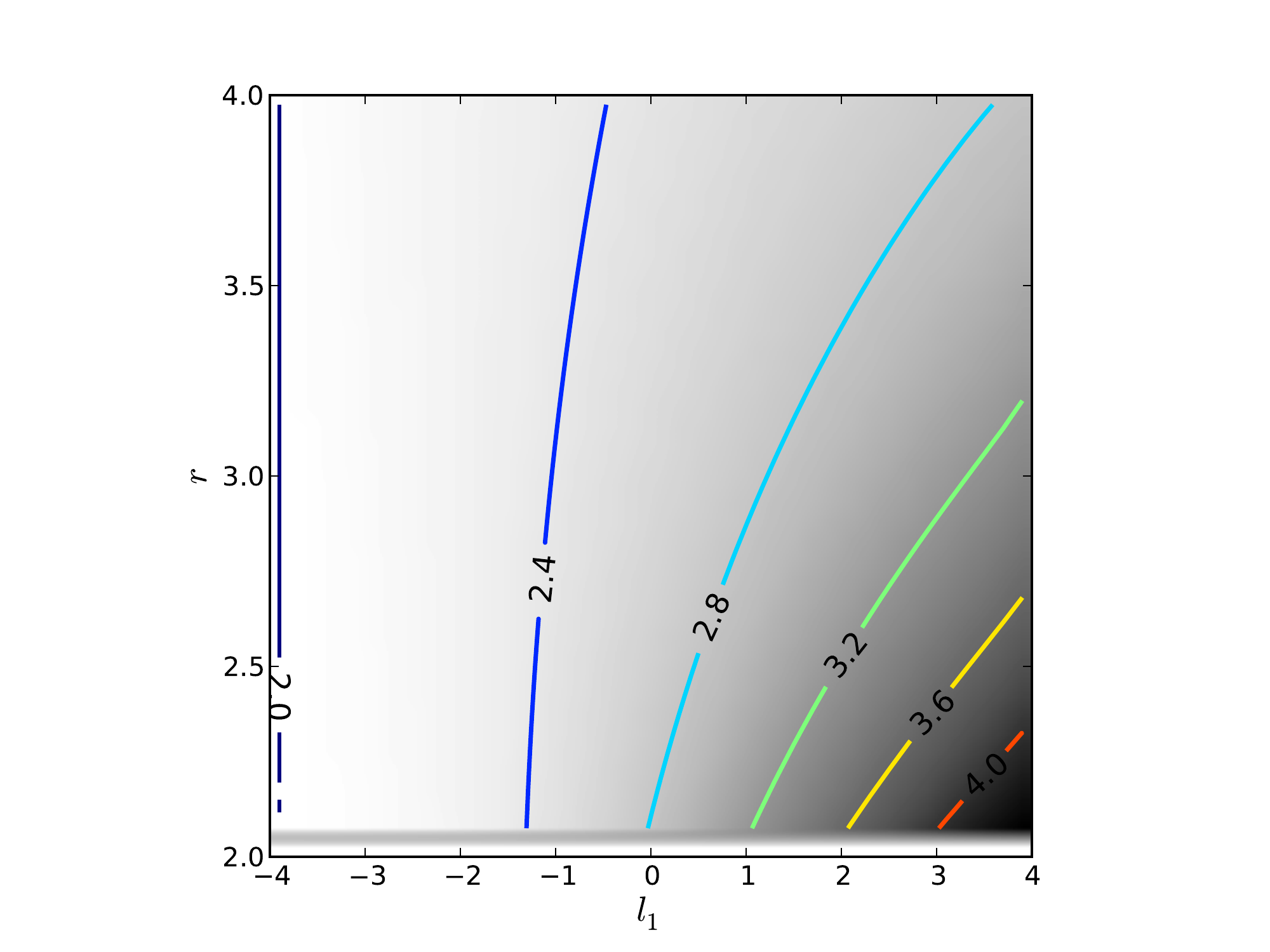}
\end{minipage}
\caption{Left: Contour plot of the CMS energy per unit DM rest mass as a function of the two DM 
angular momenta $l_1$ and $l_2$, with the collision radial coordinate being fixed at $r =r_S$. 
Right: Same as left as a function of the collision radial coordinate $r$ and the angular 
momentum of the first DM particle $l_1$, with $l_2$ fixed at its minimal value $(l_2=-4)$.}
\label{fig:ll}
\end{figure*}

Let us consider now two DM particles which, accelerated by the BH,  collide and emit photons. 
The CMS energy depends on the distance $r$ of the collision point from the BH horizon and on 
the angular momentum of the DM particles, $l_1$ and $l_2$ respectively, as 
in~\cite{Baushev:2008yz,Banados:2009pr}:
\begin{widetext}
\begin{equation}\label{eq:ecm}
\hspace*{-2cm}
E^2_{\rm CM} = 2 m^2_{B^{(1)}}  \frac{r^2 (2 r-r_S)-l_1 l_2 (r-r_S) - \sqrt{r_S r^2 -l^2_1 (r-r_S)} \sqrt{r_S r^2 - l^2_2 (r-r_S)}}{r^2 (r-r_S)}\,.
\end{equation}
\end{widetext}
To get maximal enhancement the angular momenta should be opposite in sign, as shown in 
Fig.~\ref{fig:ll} in the left panel, where the DM rest mass is fixed to unity. The enhancement in the 
CMS energy decreases fast as soon as the angular momenta decrease or as soon as the 
distance from $r_h$ increases (right panel). The Schwarzschild BH provides at most a CMS 
energy 4.5 times the DM mass. 
Such an acceleration is enough to excite the second and fourth KK levels, see 
Fig.~\ref{fig:higherres}, producing a pair of gamma ray lines. Potentially in the case of a Kerr BH 
the two gamma ray lines might become a line `forest', because of the almost infinite CMS 
energy at disposal to the DM particles.

However each emitted photon will be subject to gravitational and Doppler redshift, which in 
principle depends on where the collisions occur and in which direction the photon is emitted. 
The photons are emitted close to the BH horizon with initial energy $E_\gamma^0$ and are 
boosted with the CMS velocity $\beta$ of the annihilating DM particle pair. The gravitational 
redshift denotes the redshift in energy a photon undergoes when detected by an observer 
comoving with the source, while the Doppler shift arises by moving to the reference 
frame of a distant observer at rest. The total redshift is thus given by

\begin{equation}
E_\gamma = E^0_\gamma  \sqrt{1-\frac{r_S}{r}}\frac{\sqrt{1-\beta^2}}{1+v_{\rm tot} \cos\delta}\,,
\end{equation}
with $\delta$ being the angle at which the photon is emitted with respect to the velocity $v_{\rm tot}$ of 
the two DM particle 
system. We define a total mean redshift factor $\overline{R}_{\rm tot} $ to assess the smearing 
in the photon energy detected by a distant observer (the details are given in 
Appendix~\ref{app:BHdet}), while the spectral feature is maintained, as discussed in 
Section~\ref{sec:signew}. The total mean redshift factor  is the average over all possible 
trajectories of the emitted photons along the l.o.s. which escape the BH. The photon energy 
observed by the distant detector is then given by 
\begin{equation}\label{eq:enred}
E_\gamma = E^0_\gamma\,  \overline{R(r)}_{\rm tot} \,,
\end{equation}
while the initial injected spectrum becomes
\begin{equation}\label{eq:spred}
\frac{\der N^f_\gamma}{\der E^0_\gamma} = \frac{\der N^f_\gamma}{\der E_\gamma} \frac{\der E_\gamma}{\der E^0_\gamma} \,.
\end{equation}
Considering point sources, from Eq.~(\ref{eq:flux}) we can compute the photon flux observed at the 
Earth position from a SMBH at a distance $D$ after having integrated over the solid angle
\begin{eqnarray}
\frac{\der\Phi}{\der E_{\gamma}} & = &  \frac{1}{2 m^2_{\rm DM}} \frac{1}{D^2} \times\\\
& & \int_{r_S}^{3/2 r_S}  r^2 \rho^2(r) \left\langle\sigma v_{\rm rel} \sum_f B_f\frac{\der N_\gamma}{\der E_\gamma} \right\rangle \der r\,,\nonumber
\end{eqnarray}
where the upper limit of integration is given by the maximum value of $r$ which can lead a 
$E_{\rm CMS}$ large enough to excite at least the first the resonance, as detailed in 
Appendix~\ref{app:BHdet}. The GR effects that redshift the photons are given by Eqs.~(\ref{eq:enred}) 
and~(\ref{eq:spred}), which modify Eq.~(\ref{eq:phires}) into
\begin{eqnarray}
\label{eq:smearedflux}
 \displaystyle\frac{\der\Phi_\text{res}}{\der E_{\gamma}}  & \simeq & \frac{N_\gamma\left(\sigma v_{\rm rel}\right)_\text{res}}{4 (\overline{R(r_S)}_{\rm tot}  E^0_\gamma)^3}\times \\
& & \frac{\overline{R(r_S)}_{\rm tot} \tilde J_\text{line}(\overline{R(r_S)}_{\rm tot}  E^0_\gamma)\,\Gamma_\text{res}^2m_\text{res}^2}{(4\overline{R(r_S)^2}_{\rm tot}  (E^0_\gamma)^2-m_\text{res}^2)^2+\Gamma_\text{res}^2m_\text{res}^2}\,,\nonumber
\end{eqnarray}
with
\begin{eqnarray}\label{eq:sf1}
 \tilde J_{\rm line}(\overline{R(r_S)}_{\rm tot}  E^0_\gamma)  \equiv    \frac{1}{D^2} \frac{m_{\rm DM}^2}{t_{\rm BH}^2 (\sigma v_{\rm rel})^2
 }\times\nonumber \\
 \int_{r_S}^{3/2 r_S} \der r\, r^2  \frac{p_{\bvec r,\rm rel}(v_{\rm rel})}{v_\text{rel}/4}\,.
\end{eqnarray}
Here we used the explicit formula for the density spike, Eq.~(\ref{eq:plateau}); notice that there is no 
more $r$ dependence for the density profile as the region of interest for the signal is contained within the 
plateau region. The velocity distribution is simply given by gravitational motion around the BH and it is 
typically of the order of $c$, which we take as reference value. 
There are several astrophysical factors that affect the observed flux: 
\begin{enumerate}
\item The mean total redshift  decreases the observed photon energy. In fact, simple energy 
conservation arguments imply that the maximally observable energy for a distant observer still 
corresponds to the rest mass of the annihilating DM particles (unless very efficient Penrose 
boosting is at place);
\item In general, redshift and Doppler effects will also broaden the line signal considerably. The intrinsic 
width of the signal is much smaller than the experimental resolution, however, so we do not expect this 
to significantly affect observational prospects;
\item The distance of the SMBH tends to decrease the flux, while its Schwarzschild radius 
boosts the signal: supergiant BHs can compensate with their mass their distance and perform 
better for instance than SagA*, as it will be discussed below;
\item The time of BH formation is also relevant, as shorter $t_{BH}$  lead to 
larger fluxes;
\item The value of the annihilation cross-section in the early universe is fixed by the scaling relation in 
Eq.~(\ref{eq:scaling}) and corresponds to $\sigma v_{\rm rel} \simeq 1.3 \times 10^{-26} \rm cm^3/s$ 
for a 1.2 TeV KK particle;
\item The $p_{\bvec r,\rm rel}(v_{\rm rel})$ term, which accounts for the probability 
that the two DM particles have opposite and maximal angular momentum, in order to collide with 
a CMS energy large enough to excite the resonance. The solutions of the geodesics equations 
for massive particle would require numerical simulation. Assuming that the DM rest mass is 
negligible with respect to the CMS energy~\cite{Williams:2011uz} the geodesics can be 
approximated with those of massless objects, concluding that the infall of particles is not radial 
unless at the exact position $r=r_S$. The infall orbits are even more complicated and far from 
the radial behavior in the Kerr geometry~\cite{Bardeen:1972fi,Grib:2012iq}. In light of these 
uncertainties, we simply set $p_{\bvec r,\rm rel}(v_{\rm rel})/v_{\rm rel}/4$  to unity in our estimation of 
the photon flux near the resonances -- keeping in mind that a full treatment, which is beyond the scope 
of this work, might turn out to yield a significantly smaller value.
\end{enumerate}

 We roughly estimate the expected gamma ray flux at the Earth position for two 
 SMBHs using Eq.~(\ref{eq:smearedflux}), with $m_{B^{(1)}} = 1.2$ TeV and for the first peak produced 
 by the $H^{(2)}$ resonance, which is the dominant one.
\paragraph{SMBH in the GC} The center of the Milky Way hosts a SMBH with 
$\simeq 4.6 \times 10^6 M_{\odot}$, which corresponds to a Schwarzschild radius of about 
$ 4 \times 10^{-7}$ pc. 
For a density plateau of the order of $10^{11} \rm M_{\odot}/\rm pc^3$, derived considering a 
SMBH formation time of $10^8$ yr, the flux near the resonance is $ \Phi_{\rm res}  \sim 10^{-18}\rm 
photons/cm^2/s$, which is slightly below the reach of HESS or CTA for line searches.

\paragraph{Supergiant elliptical galaxy M87} The case of the Virgo A galaxy is more promising 
for observation, because it hosts a supergiant BH, $\rm M_{\rm BH} = 6.6 \times 10^9 M_\odot$. 
This increases considerably the Schwarzschild radius $r_S \sim 6 \times 10^{-4}$ pc, which in 
turn compensates the fact that the SMBH is much more distant than the GC, i.e. $D=16.4$ Mpc. 
In total this provides a boost of $\sim 10^3$ to the gamma ray flux with respect to the case of 
SagA*, $  \Phi_{\rm res} \sim 10^{-15}\rm photons/cm^2/s$
which might be observable by the next generation of gamma ray telescopes.

%%%%%%%%%%%%%%%%%%%%%%%%%%%%%%%%%%%%%%%%%%%%%
%%%%%%%%%%%%%%%%%%%%%%%%%%%%%%%%%%%%%%%%%%%%%
\section{Conclusions}
\label{sec:concl}

In this article, we have demonstrated that taking into account the non-vanishing velocities of
annihilating DM particles can significantly change the predictions for the  signal expected 
in indirect DM searches. This is especially true for Kaluza-Klein DM, where resonances 
naturally appear close to the CMS energy in the zero-velocity limit. For our concrete calculations  
we have focussed on a class of rather popular UED models, where the DM particle is given by 
the first KK excitation of the photon, but  the general features we have 
discussed are generic to most KK DM models. In particular, we have pointed out the remarkable 
fact that these resonances can rather generically enhance line signals significantly more than 
the continuum gamma ray spectrum from DM annihilation. This has important phenomenological 
implications for the search for TeV-scale DM candidates with upcoming instruments like CTA. 

For the model that we have chosen to investigate we have presented a systematic discussion 
of the dominant processes, for which we performed detailed higher-order computations to 
update existing results for the zero-velocity limit. This 
included hitherto neglected diagrams and a set of rather complex computations of various 
radiatively generated couplings (as explained in detail in the technical Appendices). In the 
specific case of the rather restrictive mUED model, and for typical galactic velocities,  those new 
contributions only increase the  monochromatic photon flux by $\mathcal{O}(10\%)$; for 
more general models, however, the line signal may indeed be enhanced by up to about 3 orders 
of magnitude. 

Large enhancements of the line signals can also be found in astrophysical environments where 
DM velocities larger than $\sim0.01c$ prevail. A particularly interesting place to look for line signals from 
DM annihilation are thus SMBHs like in the center of our galaxy. In this case, in fact, one may even 
encounter CMS energies several times the DM rest mass. For such a situation, we have identified a new 
'smoking gun' signature that consists of several equally spaced gamma ray lines and that would 
unequivocally point to the extra-dimensional origin of the annihilation signal. While a very rough estimate 
for the expected fluxes seems to indicate that rather favorable assumptions about the astrophysical 
environment are needed to observe such a multi-line signal, a full investigation is beyond the scope of 
this work. Given the potentially spectacular signature, however, it is certainly worthwhile to further 
explore this exciting possibility.

%%%%%%%%%%%%%%%%%%%%%%%%%%%%%%%%%%%%%%%%%%%%%
%%%%%%%%%%%%%%%%%%%%%%%%%%%%%%%%%%%%%%%%%%%%%
\begin{acknowledgments}
TB acknowledges support from the German Research Foundation (DFG) through the Emmy Noether 
grant BR 3954/1-1. The research of CA and JS has been supported at IAP by the ERC project  267117 
(DARK) hosted by Universit\'e Pierre et Marie Curie - Paris 6, PI J. Silk. JS acknowledges the support of 
the JHU by NSF grant OIA-1124403, while CA acknowledges the partial support of  the European 
Research Council through the ERC starting grant {\it WIMPs Kairos}, PI G. Bertone. MV acknowledges 
support from the Forschungs- und Wissenschaftsstiftung Hamburg through the program ``Astroparticle 
Physics with Multiple Messengers''. For the computation and numerical evaluation of the relevant 
Feynman diagrams we largely relied on {\sf FeynCalc}~\cite{Mertig:1990an} and {\sf LoopTools}~\cite{Hahn:1998yk}.
\end{acknowledgments}

\newpage

\appendix
%%%%%%%%%%%%%%%%%%%%%%%%%%%%%%%%%%%%%%%%%%%%%
%%%%%%%%%%%%%%%%%%%%%%%%%%%%%%%%%%%%%%%%%%%%%
\section{Decay widths of second-KK-level particles}\label{app:KKdec}

 As discussed in Section~\ref{sec:signew}, the way resonances decay has  phenomenological 
 consequences that are essential to our analysis.
 In this Appendix, we provide technical details about the computation of the relevant decay 
 widths of potentially resonant KK particles, c.f.~Fig.~\ref{fig:feynbbtogammax} and 
 Tab.~\ref{tab:resonances}. For concreteness, we will fix
\beq
 \label{scalefix}
  \Lambda R=5, \qquad R^{-1}=1.2\,\mathrm{TeV}
\eeq
% and alpha_em=1/128, alpha_strong=.1184 and sin^2\theta_W=0.23120
% $\mu=2/R$ as suggested in \cite{Cheng:2002iz}
whenever we state numerical results.

While tree-level couplings preserve KK number and straightforwardly follow from the SM 
Lagrangian in  5D (see, e.g., Refs.~\cite{Bringmann:2005xja,Gustafsson:2008zz} for a list of 
Feynman rules), the KK-number violating effective couplings that one encounters in resonant 
diagrams require a considerably more involved treatment.
The general formalism to obtain these effective vertices, which have to be renormalized by 
counterterms located on the brane, is detailed in Ref.~\cite{Cheng:2002iz}; it involves not only 
the calculation of radiative corrections to the vertex on the brane but also, at the same one-loop 
order, kinetic and mass mixing effects between states of different KK number. Here, we will in 
particular make use of the result that for the coupling of a generic gauge field $A_\mu$ to chiral 
SM fermions $f$, 
\beq
\mathcal{L}_\text{eff}\supset g_\text{eff}^{A\bar ff}\,A_{\mu\ a}^{(2)}\bar f^{(0)}\gamma^\nu 
T^a \frac{1\pm\gamma^5}{2} f^{(0)}\,,
\eeq
the coupling constant for the corresponding vertex in the mUED case is given by
\begin{equation}
\label{eq:effg}
g_\text{eff}^{A\bar ff}=\frac{g}{\sqrt2}\left[\frac{\bar\delta(m^2_{A^{(2)}})}{m_2^2}
-2\frac{\bar\delta(m_{f^{(2)}})}{m_2}\right]\,,
\end{equation}
where $g$ is the corresponding coupling between zero modes, $m_n\equiv n/R$ and 
$\bar\delta(m)$ refers to radiative mass corrections due to terms localized on the brane.

\subsection{$B^{(2)}$ decay}
In the mUED model the $B^{(2)}$ is the lightest of all level-2 KK particles, its mass being almost 
unaffected by radiative corrections. Kinematically, the only possible decay is directly into SM 
particles by means of KK-number violating effective vertices and its leading decay channel is 
$B^{(2)}\to\bar qq$ with a branching ratio of around 99\%  \cite{Belanger:2010yx}. 
The mass corrections of  $B^{(2)}$  and $f_{s,d}^{(1)}$ are given 
by \cite{Cheng:2002iz,Belanger:2010yx}
% \cite{Cheng:2002iz}: wrong  reference, right formulas
% \cite{Belanger:2010yx} right reference, wrong formulas
\bea
\frac{\bar\delta m^2_{B^{(2)}}}{m_2^2}&=&-\frac{g'^2}{6}\frac{\log\frac{\Lambda^2}{\mu^2}}{16\pi^2}\,,\\
\frac{\bar\delta m_{f_s^{(1)}}}{m_2}&=&\Bigg{(}\frac94Y_{f_s}^2g'^2+
3g_s^2\!-\!\frac32 y_f^2\Bigg{)}\frac{\log\frac{\Lambda^2}{\mu^2}}{16\pi^2}\,,\\
\frac{\bar\delta m_{f_d^{(1)}}}{m_2}\!&=&\! \Bigg{(}\frac94Y_{f_d}^2g'^2+\frac{27}{16}g^2
+3g_s^2-\frac34y_f^2\Bigg{)}\frac{\log\frac{\Lambda^2}{\mu^2}}{16\pi^2}\nonumber\,,\\
\eea
where $Y$ refers to the hypercharge, $g'$ [$g$] denotes the $U(1)$ [$SU(2)$] coupling 
constant and $y$ the Yukawa coupling. The term proportional to the strong coupling constant 
$g_s$ only appears for quarks.
Using Eq.~(\ref{eq:effg}), this translates into the vertex relevant for $B^{(2)}\to\bar ff$ (as 
reported in Ref.~\cite{Belanger:2010yx}) 
%misused eq. \ref{eq:effg}...)!?  
\bea
\label{B2vertex}
  \mathcal L_\text{eff}&\supset&-\bar f\gamma^\mu \left(g^L_\text{eff} \frac{1-\gamma_5}{2}+g^R_\text{eff} 
  \frac{1+\gamma_5}{2}\right)fB^{(2)}_\mu\,, \\
g^L_\text{eff} &=&\!\frac{g'Y_{f_d}}{\sqrt 2}\!\left[\frac{g'^{2}}{6}(1\!+\!27Y_{f_d}^2)+\frac{27}{8}g^2+
6g_s^2-\frac32y_f^2\right]\!\frac{\log\frac{\Lambda^2}{\mu^2}}{16\pi^2}\,,\nonumber\\ \\
g^R_\text{eff} &=&\!\frac{g'Y_{f_s}}{\sqrt 2}\left[\frac{g'^{2}}{6}(1\!+\!27Y_{f_s}^2)
+6g_s^2-3y_f^2\right]\frac{\log\frac{\Lambda^2}{\mu^2}}{16\pi^2}\,.
\eea
The decay rate then follows straight-forwardly as
\bea
\label{Bdecay}
\frac{\Gamma_{B^{(2)}\to\bar ff}}{m_{B^{(2)}}} &=& \frac1{12\pi} 
\left(1-\frac{4m_f^2}{m^2_{B^{(2)}}}\right)^\frac12\times\\
&&\times\left[\left(1+\frac{2m_f^2}{m^2_{B^{(2)}}}\right)g_V^2
 +\left(1-\frac{4m_f^2}{m^2_{B^{(2)}}}\right)g_A^2\right]\,,\nonumber
\eea
where $g_V\equiv (g^R_\text{eff}+g^L_\text{eff})/2$ and $g_A
\equiv (g^R_\text{eff}-g^L_\text{eff})/2$.
Adopting Eq.~(\ref{scalefix}), the above evaluates numerically to $g^L_\text{eff}=0.09$ and 
$g^R_\text{eff}=0.11$, implying a decay rate of $\Gamma_{B^{(2)}}\approx 0.813 \GeV$. 

When allowing for arbitrary mass splittings on the other hand, as possible in generic UED 
scenarios, $B^{(2)}$ would instead mainly decay into a pair of first-KK-level charged leptons; the 
corresponding rate is then given by
\beq
\label{Bdectree}
\frac{\Gamma_{B^{(2)}\to\bar l_s^{(1)}l_s^{(1)}}}{m_{B^{(2)}}}=
%3\times\frac13
\frac{Y_{l_s}^2g'^2}{4\pi}
\left(1-\frac{4m^2_{l_s^{(1)}}}{m^2_{B^{(2)}}}\right)^\frac12
\left(1+\frac{2m_{l_s^{(1)}}^2}{m^2_{B^{(2)}}}\right)\,.
\eeq
In this case, the existence of a tree-level rather than loop-suppressed coupling typically 
over-compensates the additional phase-space suppression, and the decay happens considerably 
faster (with $\Gamma\sim10\GeV$).

\subsection{$A_3^{(2)}$ decay}
In the mUED model, $A_3^{(2)}$ is considerable more massive than $B^{(2)}$. As a result,  
kinematics allows for six different decay channels into first-level leptons 
$\bar\ell^{(1)}_{s,d}\ell^{(1)}_{s,d}$, which  dominates over the only other possible tree-level 
decay channels into first-KK-level scalar pairs. The corresponding decay rate is described by 
Eq.~(\ref{Bdectree}), with $Yg'\to g/\sqrt{2}$. Adopting Eq.~(\ref{scalefix}), the total decay rate is 
well approximated by $\Gamma_{A_3^{(2)}}\approx70$\,GeV.

In non-minimal scenarios, mass splittings may kinematically not allow the $A_3^{(2)}$ to decay 
into first-KK-level states, but only into SM particles. Assuming that non-trivial mass and kinetic 
boundary terms are added  at the cut-off scale,
% i.e. assuming the same couplings as in mUED
 in analogy to 
Eq.~(\ref{B2vertex}), the radiative vertex that couples $A_3^{(2)}$ to SM fermion is given by
\bea
\label{A3vertex}
  \mathcal L_\text{eff}&\supset&-g^L_\text{eff} \bar f\gamma^\mu  
  \frac{1-\gamma_5}{2}fA^{(2)}_{3\mu}\,, \\
g^L_\text{eff} &=&\!\frac{g T_{3f}}{\sqrt 2}\!\left[\frac{9Y_{f_L}^2}2g'^2-\frac{33}{8}g^2+6g_s^2
-\frac32y_f^2
\right]\!\frac{\log\frac{\Lambda^2}{\mu^2}}{16\pi^2}\,,\nonumber\\
\eea
where $T_{3f}$ is the fermion weak isospin charge. In deriving this, we used again formula (\ref{eq:effg}) and also \cite{Cheng:2002iz}
\beq
\frac{\bar\delta m^2_{W^{(2)}}}{m^2_2}=\frac{15}{2}\frac{g^2}{16\pi^2}\log\frac{\Lambda^2}{\mu^2}\,.
\eeq
 The decay rate is then calculated just like in Eq.~(\ref{Bdecay}). Numerically, the total width 
 becomes $\Gamma_{A_3^{(2)}}\approx0.8$\,GeV, with branching ratios of 
 $11.2\%$ for $\bar tt$, $11.4\%$ for $\bar bb$, $18.0\%$ for other quark-antiquark pairs and 
 $0.9\%$ for every lepton pair.

\subsection{$H^{(2)}$ decay}
In the mUED model the only possible tree-level decay of the second KK-level Higgs, 
$H^{(2)}\to a_0^{(1)}B^{(1)}$,  becomes kinematically forbidden for $R^{-1}\lesssim1\TeV$. 
Therefore,  $H^{(2)}$ is also metastable and decays predominantly into top anti-top pairs due to 
a radiatively generated vertex $\mathcal L_\text{eff}\supset \,g_\text{eff}H^{(2)}\bar tt$ , 
where ($\lambda_h$ being the quartic coupling of the Higgs potential)
\cite{Belanger:2010yx}\footnote{Here, we corrected a similar formula found 
in \cite{Belanger:2010yx} by including scalar-vector-fermion loops. See 
appendix \ref{app:h2kres} for more details.}
\beq
\nonumber
g_\text{eff} = \frac{y_t}{12}\left[16g_s^2+\frac{33}{4}g^2
+\frac{23}6g'^2-9y_t^2+3\lambda_h\right]\!\frac{\log\frac{\Lambda^2}{\mu^2}}{16\pi^2}\, .
\eeq
From this, the decay rate follows as
\beq
\frac{\Gamma_{H^{(2)}\to\bar tt}}{m_{H^{(2)}}}=\frac{3g_\text{eff}^2}{8\pi} 
\left(1-\frac{4m_t^2}{m^2_{H^{(2)}}}\right)^\frac32\,.\nonumber
\eeq
The factor 3 accounts for the number of colors. Numerically, adopting Eq.~(\ref{scalefix}), we find $g_\text{eff}=0.0189$ and thus a decay rate of 
$\Gamma_{H^{(2)}}\approx 99.7 \MeV$. Note that the decay of $H^{(2)}$ in SM gauge bosons is
suppressed by a factor of roughly $\frac13(m_W/m_t)^2 \sim\mathcal{O}(0.1)$.
The decay into a $B^{(1)}$ pair, finally, is the only allowed decay channel into KK excitations in 
the mUED scenario (c.f.~Fig.~\ref{fig:atomicphysplot}). Due to the small mass splitting,
however,  this channel contributes at an even lower rate (with $\Gamma_{H^{(2)}}\approx2.4$\,MeV).

Equipping the $H^{(2)}$ with a sufficiently large mass in non-minimal scenarios, on the other 
hand, it will mainly decay into $\bar t_s^{(1)}t_d^{(1)}$ and $\bar t_d^{(1)}t_s^{(1)}$ pairs. In this 
case, one has an {\it axial} scalar coupling with
\beq
\label{geffhtt}
 g_\text{eff}=2g\frac{m_f}{m_W}\,.
\eeq
Numerically, this gives $\Gamma_{H^{(2)}}\sim 160 \GeV$, i.e.~a much faster decay than in the 
mUED scenario.

%%%%%%%%%%%%%%%%%%%%%%%%%%%%%%%%%%%%%%%%%%%%%
%%%%%%%%%%%%%%%%%%%%%%%%%%%%%%%%%%%%%%%%%%%%%
\section{Resonant LKP annihilation amplitudes}
\label{app:KKann}
In this Appendix, we provide technical details about the computation of the full amplitudes 
that describe the resonant annihilation of LKP pairs.
The Feynman diagrams of Fig.~\ref{fig:feynbbtogammax} can compactly be written as
\begin{eqnarray}
i\mathcal A_\text{V}&=&\epsilon_1^\mu\epsilon_2^\nu L^{\ \rho}_{\text{V}\mu\nu}\frac{-\eta_{\rho\sigma}+
\frac{P_\rho P_\sigma}{M^2_\text{V}}}{s-M^2_\text{V}+iM_\text{V}\Gamma_\text{V}}R^{\ \sigma}_{\text{V}\alpha}\epsilon_\gamma^\alpha\ ,\\
i\mathcal A_\text{S} &=& \epsilon_1^\mu\epsilon_2^\nu L_{\text{S}\mu\nu}\frac{1}{s-M^2_\text{S}
+iM_\text{S}\Gamma_\text{S}}R_{\text{S}\alpha\beta}\epsilon_\gamma^\alpha\epsilon_Z^\beta\ ,
\end{eqnarray}
where $V$ and $S$ stand for vector ($B^{(2)}$, $A_3^{(2)}$) and scalar ($H^{(2)}$, $a_0^{(2)}$) 
resonances respectively. The tensors $L_{V,S}$, $R_{V,S}$ encode therefore the physical 
information of the left and right blobs in each diagram of Fig. \ref{fig:feynbbtogammax}. In the 
following, we will focus our discussion of these tensors in a final-state-to-final-state basis.

\subsection{$B^{(1)}B^{(1)}\to\gamma\gamma$}
\label{appss:h2res}
The relevant Feynman diagram that contributes to this process is
\begin{center}
{}\hspace{1cm}{} \includegraphics[width=.8\linewidth]{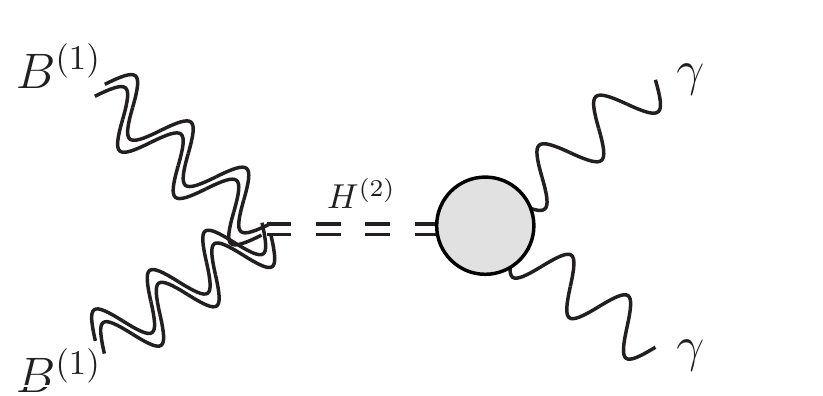}
\end{center}
Here, $L_{\mu\nu}=ig_{B^{(1)}B^{(1)}H^{(2)}}\eta_{\mu\nu}=i(g'^2v/2)\eta_{\mu\nu}$, where $v$ 
is the vacuum expectation value of the Higgs field and the blob on the right-hand-side 
represents the superposition of several triangle diagrams, the leading ones being
\begin{center}
\includegraphics[width=.4\linewidth]{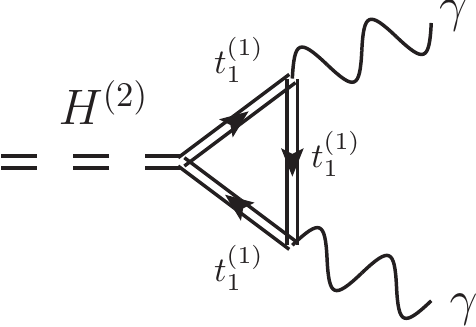}\ \includegraphics[width=.4\linewidth]{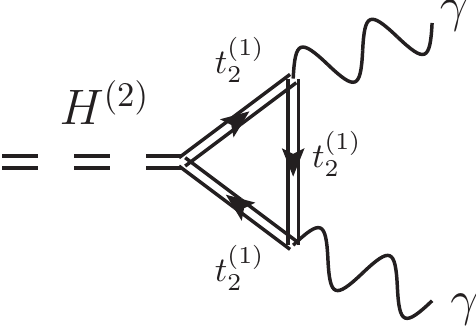}
\end{center}
Decomposing the corresponding loop-integrals in terms of Passarino-Veltman 
functions \cite{Passarino:1978jh} yields (in the limit where both KK top quarks have the same 
mass) 
\bea
\label{Rhgg}
R^{\gamma\gamma}_{H^{(2)}\alpha\beta} &=&-\frac{\alpha_\mathrm{em}Q_t^2}{\pi}
\frac{4 igm_t \sin2\alpha_t^{(1)}}{m_W}\frac{m_{t^{(1)}}}{s}\\
&&\Bigg(
\Big[2\!-\!(s\!-\!4m_{t^{(1)}}^2)\textrm{C}_0(0,0,s,m_{t^{(1)}}^2,m_{t^{(1)}}^2,m_{t^{(1)}}^2\!)\Big]\nonumber\\
&&\quad\times\left[s\eta_{\alpha\beta}\!-\!\!2k_{1\alpha}k_{2\beta}-\!\!2k_{1\beta}k_{2\alpha}\right]\nonumber\\
&&-4\Big[2\textrm{B}_0(s,m_{t^{(1)}}^2,m_{t^{(1)}}^2)-2\textrm{B}_0(0,m_{t^{(1)}}^2,m_{t^{(1)}}^2)\nonumber\\
&&\quad +s\,\textrm{C}_0(0,0,s,m_{t^{(1)}}^2,m_{t^{(1)}}^2,m_{t^{(1)}}^2)\Big] k_{1\alpha}k_{2\beta}\Bigg)\,.\nonumber
\eea
Here, $k_1$ and $k_2$ are the outgoing momenta and $s=(k_1+k_2)^2$, 
$\alpha_em\approx1/128$ is the fine-structure constant at the TeV scale and $Q_t=2/3$ is the 
charge of the top-quark. The angle $\alpha^{(1)}$ describes the mixing between first KK-level 
flavour and mass eigenstates; in the mUED case, this is only significantly different from zero for 
the case of top quarks (with $\alpha_t^{(1)}\approx 0.071$).
Notice that $R_{H^{(2)}\alpha\beta}$ manifestly satisfies the Ward identities both here and for 
the amplitudes presented further down, namely $k_1^\alpha R_{H^{(2)}\alpha\beta}=k_2^\beta R_{H^{(2)}\alpha\beta}=0$.

Concerning a possible contribution from the $a_0^{(2)}$ resonance, we note that 
$ig_{B^{(1)}B^{(1)}a_0^{(2)}}=0$ at tree level and, more importantly,  
$\Gamma_{a_0^{(2)}\to\gamma\gamma,\gamma Z}=0$. The latter can be traced back to the 
absence of anomalous three-gauge-boson couplings in the full theory (recall that $a_0$ 
contains the higher-dimensional component of the $Z$ boson).\vspace{0.3cm}
% technically: $a_0^{(2)}$ doesn't couple singlets (doublets) with singlets (doublets) at all (contrary to $H^{(2)}$)

\subsection{$B^{(1)}B^{(1)}\to Z\gamma$}
Most of the conclusions from the last section can also be drawn for this process. The main 
difference is that the  $Z$ boson is massive and has not only vector, but also axial-vector 
couplings. The following two diagrams thus need in principle to be added for the computation of 
$R_{\alpha\beta}$ in this case, but cancel in the limit of $m_{t^{(1)}_1}=m_{t^{(1)}_2}$ :\vspace{0.01cm}
\begin{center}
\includegraphics[width=.4\linewidth]{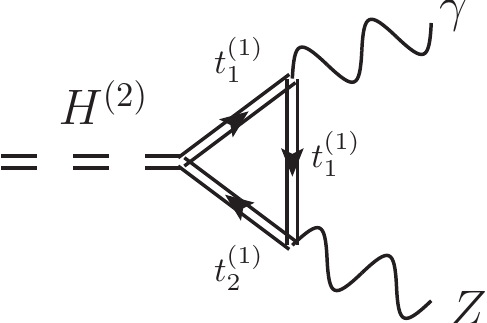}\ \includegraphics[width=.4\linewidth]{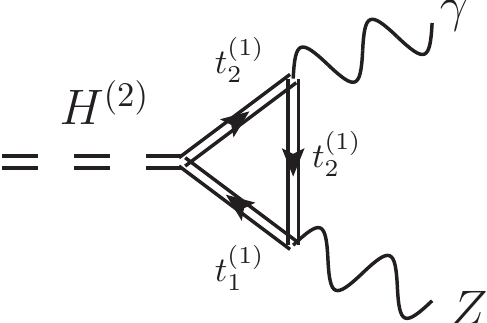}
\end{center}
In total, we find
%k_1:photon, k_2: Z
\begin{widetext}
\begin{eqnarray}
\label{RhgZ}
R^{\gamma Z}_{H^{(2)}\alpha\beta}&=&
-\frac{egQ_t}{\pi^2\cos\theta_W} \frac{igm_t}{m_W}\frac{m_{t^{(1)}}\sin2\alpha_t^{(1)}}{s-m_Z^2}
\left(
B\,[(s\!-\!m_Z^2)\eta_{\alpha\beta}-2k_{1\beta}k_{2\alpha}]
+C\,[k_{2\beta}\!-\!\frac{2m_Z^2}{s-m_Z^2}k_{1\beta}]k_{1\alpha}
\right)\nonumber\\
\textrm{with}\\
B&=&\left(Y_{t_L}\sin^2\theta_W-\frac14\cos2\theta_W\right)[2-(s-m_Z^2-4m_{t^{(1)}}^2)\textrm{C}_0(0,m_Z^2,s,m_{t^{(1)}}^2,m_{t^{(1)}}^2,m_{t^{(1)}}^2)- \\
& &-\frac{2m_Z^2}{s-m_Z^2}[\textrm{B}_0(s,m_{t^{(1)}}^2,m_{t^{(1)}}^2)-\textrm{B}_0(m_Z^2,m_{t^{(1)}}^2,m_{t^{(1)}}^2)]] \nonumber\\
 C&=&\!\!2\!\left(Y_{t_L}\sin^2\theta_W\!-\!\frac14\cos2\theta_W\right)[2+(s+m_Z^2+4m_{t^{(1)}}^2)\textrm{C}_0(0,m_Z^2,s,m_{t^{(1)}},m_{t^{(1)}},m_{t^{(1)}}) \\
& &+2\frac{2s+m_Z^2}{s-m_Z^2}\textrm{B}_0(s,m_{t^{(1)}}^2,m_{t^{(1)}}^2)-2\frac{s+2m_Z^2}{s-m_Z^2}\textrm{B}_0(m_Z^2,m_{t^{(1)}}^2,m_{t^{(1)}}^2)-2\textrm{B}_0(0,m_{t^{(1)}}^2,m_{t^{(1)}}^2)]\ . 
\end{eqnarray}
\end{widetext}

\subsection{$B^{(1)}B^{(1)}\to H\gamma$}
In this case, the relevant Feynman diagrams are given by
\begin{center}
\includegraphics[width=.6\linewidth]{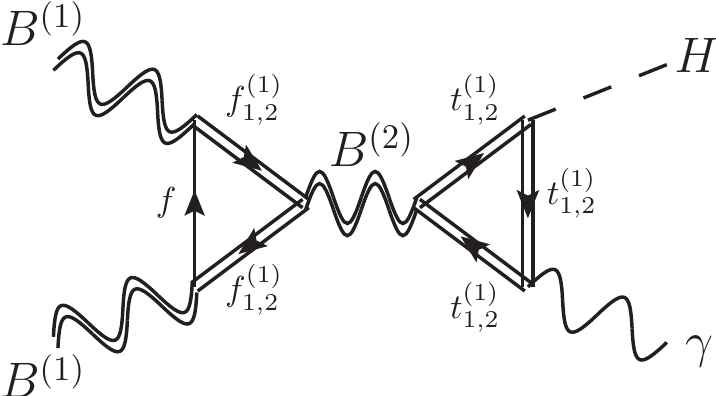}
\end{center}
where the vector resonance $B^{(2)}$ can also be interchanged with $A_3^{(2)}$. The reason 
that top quark contributions dominate in the tensor  $R_{V}$ is, as in the cases discussed so far, 
simply given by the presence of a Yukawa. For $L_{V}$, on the other hand, this  follows from 
anomaly cancellation in the SM: 
$\sum_fL^\rho_{f\mu\nu}|_{m_f\to0}\propto\sum_f(Y_R^3+Y_L^3)=0$ implies that $L^\rho_{\mu\nu}=\sum_fL^\rho_{f\mu\nu}\simeq L^\rho_{\text{top}\,\mu\nu}- L^\rho_{\text{top}\,\mu\nu}|_{m_t\to0}$ 
(with the sum running over all SM fermions $f$). While this results in an expression for 
$L^\rho_{\mu\nu}$ that is too lengthy to be displayed here, the tensor 
$R^{\gamma H}_{B^{(2)}\alpha\beta}$ takes a very similar form as in the previous case:
%convention: Y_t_L=1/6, Y_t_R=2/3
\begin{widetext}
\begin{eqnarray}
\label{RV}
 R^{\gamma H}_{B^{(2)}\alpha\beta}&=&-\frac{eg'Q_t (Y_{t_L}+Y_{t_R})}{2\pi^2}\frac{igm_t}{m_W}\frac{m_{t^{(1)}}\sin2\alpha_t^{(1)}}{(s-m_H^2)^3}
\Big{(}B[(s-m_H^2)\eta_{\alpha\beta}-2k_{1\beta}k_{2\alpha}]+[C_2k_{2\beta}-C_1k_{1\beta}]k_{1\alpha}\Big{)}\\
\textrm{with}\nonumber\\
B&=&(s-m_H^2)([2+(s-m_H^2+4m_{t^{(1)}}^2)\textrm{C}_0(0,m_H^2,s,m_{t^{(1)}}^2,m_{t^{(1)}}^2,m_{t^{(1)}}^2)](s-m_H^2)\\
&&-2s[\textrm{B}_0(s,m_{t^{(1)}}^2,m_{t^{(1)}}^2)-\textrm{B}_0(m_H^2,m_{t^{(1)}}^2,m_{t^{(1)}}^2)])\nonumber\\
C_2&=&\!\!-2(s-m_H^2)[(s-m_H^2)[2+(s+m_H^2+4m_{t^{(1)}}^2)\textrm{C}_0(0,m_H^2,s,m_{t^{(1)}},m_{t^{(1)}},m_{t^{(1)}})]+2(s+2m_H^2)[\textrm{B}_0(s,m_{t^{(1)}}^2,m_{t^{(1)}}^2)\nonumber\\
&&-\textrm{B}_0(m_H^2,m_{t^{(1)}}^2,m_{t^{(1)}}^2)]+2(s-m_H^2)[\textrm{B}_0(s,m_{t^{(1)}}^2,m_{t^{(1)}}^2)-\textrm{B}_0(,m_{t^{(1)}}^2,m_{t^{(1)}}^2)]\nonumber\\
C_1&=&4[m_H^2(s-m_H^2)[2+(s+2m_{t^{(1)}}^2)\textrm{C}_0(0,m_H^2,s,m_{t^{(1)}},m_{t^{(1)}},m_{t^{(1)}})]+2m_H^2(2s+m_H^2)[\textrm{B}_0(s,m_{t^{(1)}}^2,m_{t^{(1)}}^2)\nonumber\\
&&-\textrm{B}_0(m_H^2,m_{t^{(1)}}^2,m_{t^{(1)}}^2)]+(s^2-m_H^4)[\textrm{B}_0(s,m_{t^{(1)}}^2,m_{t^{(1)}}^2)-\textrm{B}_0(,m_{t^{(1)}}^2,m_{t^{(1)}}^2)]\ .\nonumber
\end{eqnarray}
\end{widetext}
For $R^{\gamma H}_{A_3^{(2)}\alpha\beta}$, one simply needs to replace $\sin2\alpha^{(1)}_t(Y_{t_L}+Y_{t_R})$  in Eq.~(\ref{RV}) with $(1/4)\sin4\alpha^{(1)}_t$.

%%%%%%%%%%%%%%%%%%%%%%%%%%%%%%%%%%%%%%%%%%%%%%%
%%%%%%%%%%%%%%%%%%%%%%%%%%%%%%%%%%%%%%%%%%%%%%%

\section{\boldmath{$H^{(2k)}$} resonances}
\label{app:h2kres}

Similar to $H^{(2)}$, higher order scalar modes decay via KK-number violating processes in 
minimal scenarios. However, these decays occur with non-trivial branching ratios because, as 
the KK-mode increases, more and more final states start to become 
kinematically available. Therefore a careful analysis of the effective brane vertices describing such 
decays is needed. Fortunately, we can derive all of them from the master 5D vertex
and from the kinetic and mass radiative mixing terms \cite{Cheng:2002iz}. In Feynman gauge,
the Vertex reads
\begin{align}
\nonumber
\delta&\mathcal L_\text{eff} = L^{3/2}\left(\frac{\delta(x_5)+\delta(x_5-L)}2\right)\frac{y_t}{\sqrt 2}\frac1{64\pi^2}\log\frac{\Lambda^2}{\mu^2}\times\\
\label{eq:ttHeffvertex}
{}&\times\left[f_RH\bar t_d\frac{1+\gamma_5}2t_s+f_LH\bar t_d\frac{1-\gamma_5}2t_s+\text{h.c.}\right]\,,
\end{align}
where $x_5$ is the fifth space coordinate, $L=\pi R$ is the length separating the orbifold 
fixed points and $f_{L,R}$ are given by
\begin{eqnarray*}
f_R &=& 8g_s^2-\frac32g^2-\frac16g'^2\,,\\
% {} &=&6\left(\frac43\right)g_s^2\!+\!\left(6Y_{t_R}Y_{t_L}\!\!+2(|Y_{t_R}|\!+\!|Y_{t_L}|)\frac12\right)g'^2\!\!+\\
% {} &{}&+2\left(\frac34\right)\!g^2\ ,\\
f_L &=& -2y_t(y_t+y_b)\approx-2y_t^2\,,
\end{eqnarray*}
The mixing terms, on the other hand, are given by
\begin{align}
\nonumber
\delta\mathcal L\supset L\left(\frac{\delta(x_5)+\delta(x_5-L)}2\right)&\frac1{64\pi^2}\log\frac{\Lambda^2}{\mu^2}\times\\
\times[b^{s,d}_1\bar t_{s,d}i\cancel\partial t_{s,d;+,-}+b^{s,d}_2(&\bar t_{s,d}\overleftarrow\partial_5t_{s,d;+,-}+\\
\nonumber
+\bar t_{s,d;+,-}\partial_5t_{s,d})+c_1\frac12(\partial_\mu H)^2&+c_2\frac12 H\partial_5^2 H]\,,
\end{align}
where the fields $(1\pm\gamma_5)t/2$ are represented as $t_\pm$, and $b^{s,d}_{1,2}$ and $c_{1,2}$ are
given by \cite{Cheng:2002iz}
\bea
b_1^s&=&\frac43g_s^2+Y_{t_R}^2g'^2+2y_t^2\,,\\
b_1^d&=&\frac43g_s^2+\frac34g^2+Y_{t_L}^2g'^2+y_t^2\,,\\
b_2^s&=&5\left(\frac43g_s^2+Y_{t_R}^2g'^2\right)-2y_t^2\,,\\
b_2^d&=&5\left(\frac43g_s^2+\frac34g^2+Y_{t_L}^2g'^2\right)-y_t^2\,,\\
c_1&=&-g'^2-2g^2\,,\\
c_2&=&\frac12g'^2+g^2-2\lambda_h\,.
\eea

\subsection{\boldmath{$H^{(4)}$} decay}
The computation of decay rates for $H^{(2k)}$ in the minimal scenario follows from similar 
procedures as the corresponding calculation for $H^{(4)}$. Here we therefore compute this 
decay rate as an example.

Notice that the main difference between the life-times of $H^{(2)}$ and $H^{(4)}$ is that whereas 
the former decays with branching ratio $\sim 1$ into top quark-antiquark pairs, the latter can 
decay into several states with comparable branching ratios. Namely, 
$H^{(4)}\to\bar t_s^{(1)}t_d^{(1)}(\bar t_d^{(1)}t_s^{(1)})$, $\bar t_{s,d}^{(2)}t^{(0)}$ ($\bar t^{(0)}t_{s,d}^{(2)}$) and $\bar tt$ 
turn out to be the 7 dominant decay channels, where for instance 
the decays $H^{(4)}\to\bar t_{s,d}^{(1)}t_{d,s}^{(1)}$ are induced by the effective vertex
\begin{align*}
\delta\mathcal L\supset\frac{y_t}{64\pi^2}\log\frac{\Lambda^2}{\mu^2}&H^{(4)}\Bigg{[}\Bigg{(}\!\!\frac{154}{9}g_s^2+\frac{41}{16}g^2+\frac{697}{432}g'^2-\\
-\frac{23}{4}y_t^2+\frac83\lambda_h\Bigg{)}\bar t^{(1)}_s&t^{(1)}_d+\Bigg{(}\frac{69}{9}g_s^2-\frac{29}{16}g^2-\frac{157}{432}g'^2+\\
+\frac{3}{4}y_t^2&\Bigg{)}\bar t^{(1)}_s\gamma_5t^{(1)}_d+\text{h.c.}\Bigg{]}\,,
\end{align*}
which is obtained by decomposing (\ref{eq:ttHeffvertex}) in terms of the KK modes and including 
the kinetic and mass mixing terms with $t_{s,d}^{(3)}$, $t_{s,d}^{(5)}$, $H^{(0)}$ and $H^{(2)}$. 
The numerical value for the total decay rate of $H^{(4)}$ amounts to 
$\Gamma_{H^{(4)}}=3.1\GeV$ when $\Lambda=5/R$.

\subsection{\boldmath{$B^{(1)}B^{(1)}H^{(2k)}$} effective vertices}
\label{app:effvert}
In section \ref{appss:h2res} we exploited the fact that EW-symmetry breaking provides us with a 
tree-level $B^{(1)}B^{(1)}H^{(2)}$ coupling when obtaining the 
$B^{(1)}B^{(1)}\to\gamma\gamma(\gamma Z)$ annihilation rates. Such a coupling does not 
exist for, say, $B^{(1)}B^{(1)}H^{(4)}$ in the classical theory since it violates KK-number 
symmetry. However, the same arguments from the previous section apply here and we find 
couplings of this kind at the loop-quantum level which are localized at the fixed points of the 
orbifold. 

To obtain these effective vertices, we shall just as previously consider the master 5D radiative 
terms
\begin{equation}
\delta\mathcal L\supset L\left(\frac{\delta(x_5)+\delta(x_5-L)}2\right)\frac{g'^2v}2\frac{f_S}{64\pi^2}\log\frac{\Lambda^2}{\mu^2}HB_\mu B^\mu
\label{eq:BBHeffvertex}
\end{equation}
and the kinetic and mass mixing terms
\begin{align}
\nonumber
\delta\mathcal L\supset L\left(\frac{\delta(x_5)+\delta(x_5-L)}2\right)\frac1{64\pi^2}&\log\frac{\Lambda^2}{\mu^2}[-a^B_1\frac14B_{\mu\nu}B^{\mu\nu}\\
-a^B_2\frac12(B_\mu\partial^2_5B^\mu)+c_1\frac12(\partial_\mu H)^2+&c_2\frac12 H\partial_5^2 H]\,.
\end{align}

\begin{figure}
 \includegraphics[width=\linewidth]{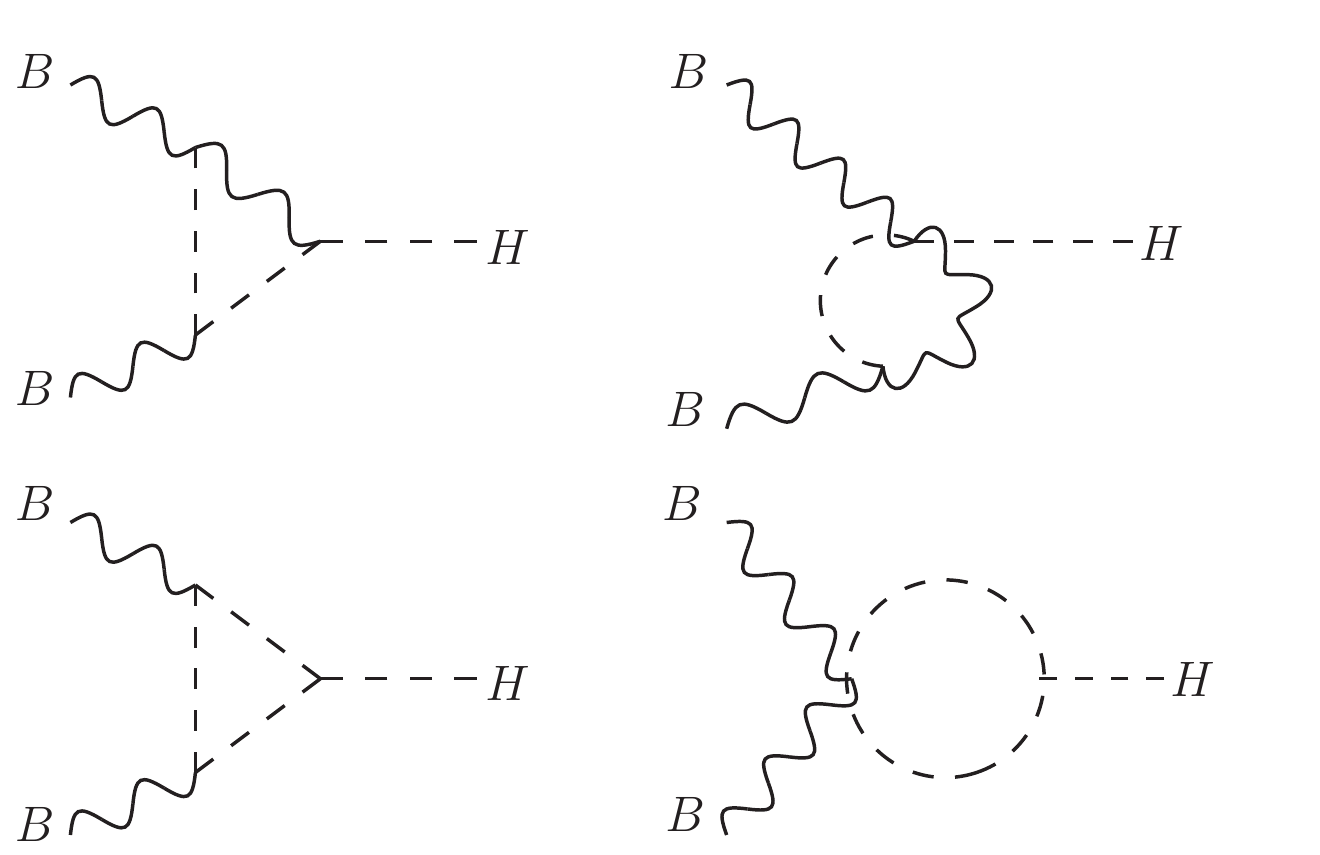}
\caption{\label{fig:BBHeffvertex}Divergent Feynman diagrams participating in eq. 
(\ref{eq:BBHeffvertex}). Additional diagrams are obtained by charge conjugation or leg 
exchange of the initial $B$'s. Particles running on the loops include (correspondingly) all vector 
bosons and all scalars. Notice that there are no fermion loops since H couples to mixed doublet 
and singlets.}
\end{figure}

The coefficients in the previous expression have already been computed in Ref.~\cite{Cheng:2002iz}, 
while the coefficient $f_S$ in (\ref{eq:BBHeffvertex}) can easily be computed by isolating the 
divergent terms of the Feynman diagrams in Fig. \ref{fig:BBHeffvertex} and adding them up. In 
the Feynman gauge the result reads
\beq
f_S=\frac34g'^2+\frac94g^2+12\lambda_h\ .
\eeq

To check the correctness of this result, one can obtain the corresponding effective vertices and 
mixing terms for the $A_3$ field -- which must be done with care due to additional types of Feynman 
diagrams (ghosts, W loops, etc)  -- and verify that terms like 
$A^{(0)}_\mu Z^{(0)\mu} H^{(2k)}$ or $A_\mu^{(0)} A^{(0)\mu} H^{(2k)}$ do not exist as required 
by gauge invariance ($A_\mu$ represents the photon field). 
%We leave that as an excercise to the reader.

%%%%%%%%%%%%%%%%%%%%%%%%%%%%%%%%%%%%%%%%%%%%%%%
%%%%%%%%%%%%%%%%%%%%%%%%%%%%%%%%%%%%%%%%%%%%%%%

\section{Details on the photon flux from DM collisions in Schwarzschild BHs}\label{app:BHdet}
The derivation and solutions of photon geodesics in the Schwarzschild metric can be found, e.g. 
in~\cite{Misner:1974qy}. Here we discuss the case relevant for the gamma ray line signature we 
present in this work.

A non rotating BH is described by the Schwarzschild metric, which in spherical coordinates is 
\begin{eqnarray}\label{eq:sbhmertic}
\der s^2  & = & -\left(1-\frac{r_S}{r}\right) \der t^2 + \left(1-\frac{r_S}{r}\right)^{-1} \der r^2 + \nonumber \\
& & r^2 \left(\der \theta^2 + \sin^2\theta \der \varphi^2 \right)\,,
\end{eqnarray}
where we have set $c=G=1$ and $r_S=2M_{\rm BH}$ is the Schwarzschild radius. The four-
velocity of a massive object is $u^\mu = \der x^\mu/\der \tau = (u^t,u^r,0,u^\varphi)$ (when 
referring to the three- component velocity of a DM particle, $v$ denotes the module of the 
velocity, while $v_{\rm rel}$ and $v_{\rm tan} $ denote the radial and tangential velocities). We 
indicate with $k^\mu$ the photon four-velocity.

As usual, the geodesics are defined as:
\begin{equation}\label{eq:geods}
\frac{\der^2 x^\alpha}{\der \tau} + \Gamma^{\alpha}_{\beta\gamma} \frac{\der x^\beta}{\der \tau}\frac{\der x^\gamma}{\der \tau}=0\,,
\end{equation}
with $\Gamma_{\alpha\beta}^\gamma$ the Christoffel symbols and $\tau$ the proper time 
(replaced with the affine parameter $\lambda$ for massless particles). 

The three-velocity components and the module of the velocity of one DM particle, from 
Eq.~(\ref{eq:sbhmertic}) and~(\ref{eq:geods}), are 
\begin{eqnarray}
 v_{\rm tot} & = &  \frac{\sqrt{r^2 r_S + r_S l^2 - r l^2}}{r\sqrt{r}}\,,\nonumber \\
 v_{\rm rel} & = & \frac{l}{r}\sqrt{1-\frac{r_S}{r}}\,,\nonumber\\
 v & = & \sqrt{\frac{r_S}{r}}\,.
\end{eqnarray}
The $v_{\rm tot} $ component, which is the velocity of the collided DM system, shows that the maximum 
allowed angular momentum for a particle falling into the BH is $|l|=4$, otherwise $\der r/\der t$ 
has a turning point before reaching the horizon. This demonstrates Eq.~(\ref{eq:ecm}) and is 
shown in Fig.~\ref{fig:ll}. 

Let us consider now the photons, with both radial and angular motion, emitted by the DM 
system. Since both energy and angular momentum are conserved but $\gamma$'s are 
massless it is useful to introduce the impact parameter $b = L/E$. The photon geodesics are 
given by
\begin{eqnarray}
 \frac{\der t}{\der \lambda}&  = & \frac{1}{b}\left(1-\frac{r_S}{r}\right)^{-1}\,,\nonumber \\
 \frac{\der r}{\der \lambda} & =  &\pm \frac{1}{b}\sqrt{1-\frac{b^2}{B^2(r)}}\,,\nonumber \\
 \frac{\der \varphi}{\der \lambda} & = & \frac{1}{r^2}\,,
\end{eqnarray}
where $B^{-2}(r) = 1/r^2 \left(1-r_S/r\right)$. The three-velocity components of the photons are
\begin{equation}
 k_r  =  \sqrt{1-\frac{b^2}{B^2(r)}}\, \, \, \, {\rm and} \, \, \, \, k_t   =  \frac{b}{B}\,,
\end{equation}
such that $k_r^2 + k_t^2 =1$. Notice that only photons satisfying the following conditions can 
escape from the BH and reach far observers:
\begin{eqnarray} \label{eq:cond}
(1) \,\,  r< \frac{3}{2}\, r_{S}\,,   k_r > 0\,, & \sin\delta < \frac{3\sqrt{3}}{2}\,  r_S (B(r))^{-1}\,,\label{eq:caseok}\\
( 2)\, \,   r > \frac{3}{2}\,  r_S \,,  k_r > 0\,, &  \label{eq:no1}\\
( 3) \, \,  r< \frac{3}{2}\,  r_S \,,   k_r < 0\,, & \sin\delta > \frac{3 \sqrt{3}}{2}\,  r_S (B(r))^{-1}\,,\label{eq:no2}
\end{eqnarray}
where $\delta \equiv \arccos k_r \equiv \arcsin k_t$. Since the Schwarzschild BH can only 
provide a significant enhancement of CMS energy close to $r=r_S$, only Eq.~(\ref{eq:caseok}) 
is relevant. Notice that if the photon is emitted exactly at $r=r_S$, it has only a radial trajectory 
and the escape condition does not depend on its energy but only on the position $r$.

The initial conditions for the photon emitted from DM annihilation are given by 
Eq.~(\ref{eq:ecm}) and by its velocity $\beta = v_{\rm rel}$, equivalent to the relative velocity of 
the DM system. Let us first assume an observer which is comoving with the center of mass 
energy of the collision, hence stationary with respect to the collided system of two DM particles 
so that  $u^t =  1/\sqrt{(1-r_S/r)}$.
The photon energy $E_\gamma$ observed far away by the comoving observer is given by the 
gravitational redshift:
\begin{equation}
E_\gamma = E^0_\gamma \sqrt{1-\frac{r_S}{r}} \,.
\end{equation}

We can then consider a stationary observer very far away from the BH, that sees the center of 
mass frame moving with velocity $\beta$, which gives the doppler shift effect added on top of 
the gravitational redshift. The frequency of the observed photon is
\begin{equation}
\omega= k^\alpha u_\alpha  = g_{\alpha\beta} u^\alpha k^\beta  = g_{tt} u^t k^t + g_{rr} u^r k^r \,,
\end{equation}
where the $g_{\alpha\beta}$ are the components of the Schwarzschild metric. Equivalently, the 
observed photon energy is:
\begin{equation}\label{eq:final}
E_\gamma = E^0_\gamma \sqrt{1-\frac{r_S}{r}}\, \frac{\sqrt{1-\beta^2}}{1+\beta}\,.
\end{equation}
More generally the Doppler shift can be a function of the angle $\delta$ (defined in 
Eqs.~(\ref{eq:caseok} -~\ref{eq:no2})) between the emitted photon and the velocity of the source 
term
\begin{equation}
E_\gamma = E^0_\gamma  \sqrt{1-\frac{r_S}{r}}\frac{\sqrt{1-\beta^2}}{1+v_{\rm tot} \cos\delta}\,.
\end{equation}
Note that the Doppler factor and the gravitational redshift factorize.

To account for all photons that can actually escape from the BH and reach the far observer 
along the line of sight on a small angle cone, we define a mean total redshift as
\begin{eqnarray}
\overline{R}_{\rm tot} & = & \frac{1}{4\pi}\int(2\pi\der\delta\sin\delta)\,  \sqrt{1-\frac{r_S}{r}}\frac{\sqrt{1-\beta^2}}{1+v_{\rm tot} \cos\delta}\, \nonumber\\
& & \times \Theta(\frac{3 \sqrt{3}}{2} \frac{r_S}{r}\sqrt{1-\frac{r_S}{r}}-\sin\delta )\,,
\end{eqnarray}
where the $\Theta$ function satisfies Eq.~(\ref{eq:caseok}). This 
can be rewritten as a function of $\cos\delta = x$, 
\begin{eqnarray}
\overline{R}_{\rm tot} & = & \frac12\sqrt{1-\frac{r_S}{r}} \sqrt{1-\beta^2 }\int_{x_{\rm min}}^1 \der x \frac{1}{1+ v_{\rm tot }x}\\
{} &=& \frac1{2v_{\rm tot}}\sqrt{1-\frac{r_S}{r}} \sqrt{1-\beta^2 }\log\left(\frac{1+v_{\rm tot}}{1+v_{\rm tot}x_{\rm min}}\right)\nonumber\,,
\end{eqnarray}
where $x_{\rm min}$ is determined by the Heavyside function.
The mean redshift factor enters in the photon flux emitted by the BH, Eq.~(\ref{eq:smearedflux}), and acts as a smearing factor.

\bibliography{biblio}
\end{document}